\documentclass{article}
\usepackage{feyn}
\usepackage{bm}
\usepackage{bookstyle}
\usepackage{graphicx}
\usepackage{epsfig}

\begin{document}

\author{Walter D. Goldberger}
\address{Department of Physics, Yale University, New Haven, CT 06520}
\title{Les Houches Lectures on Effective Field Theories and Gravitational Radiation}
\photo{}

\frontmatter
\maketitle
\mainmatter

\begin{abstract}
These lectures give an overview of the uses of effective field theories in describing gravitational radiation sources for LIGO or LISA.   The first lecture reviews some of the standard ideas of effective field theory (decoupling, matching, power counting) mostly in the context of a simple toy model.   The second lecture sets up the problem of calculating gravitational wave emission from non-relativistic binary stars by constructing a tower of effective theories that separately describe each scale in the problem:   the internal size of each binary constituent, the orbital separation, and the wavelength of radiated gravitons.

%
 \end{abstract}

\section{Lecture I}

\subsection{Introduction and motivation}
\label{sec:motivation}

These lectures describe the uses of effective field theory (EFT) methods to solve problems in gravitational wave physics.   Many of the signals relevant to gravitational wave experiments such as LIGO~\cite{LIGO}, VIRGO~\cite{VIRGO} and the planned LISA~\cite{LISA} correspond to astrophysical sources whose evolution involve a number of distinctly separated length scales.   In order to compute signal templates that capture the physics accurately, it is important to systematically account for effects arising at all these different scales.   
It is exactly this sort of problem that is best treated by EFT methods, analogous to the EFTs constructed to unravel multiple scale problems in high energy physics or condensed matter.

For the sake of concreteness, I will focus in these lectures on the EFT formulation of the slow  ``inspiral'' phase of compact binary stars (that is, with neutron star (NS) or black hole (BH) constituents).   A more comprehensive review of gravitational wave sources and phenomenology can be found in the lectures by A. Buonnano at this school (also see, e.g., ref.~\cite{GWrev}).   The inspiral phase plays an important role in gravitational wave physics, and corresponds to the period in the evolution of the binary in which the system is non-relativistic, the bound orbit slowly decaying due to the emission of gravitational radiation.    

An appealing feature of the binary inspiral phase is that, theoretically, it is a very tractable problem.    In principle, the dynamics is calculable as a perturbative expansion of the Einstein equations in the parameter $v\ll 1$, a typical three-velocity.\footnote{Even though we will be doing classical physics for most of these lectures, I use particle physics units $c=\hbar=1$ throughout.}    To get a handle on the scales involved, it is worthwhile to calculate some of the features of the binary inspiral at the order of magnitude level.    Let's consider inspiral events seen by LIGO, which for illustration we will assume operates in the frequency band between 10 Hz and 1 kHz.   At such frequencies, the relevant inspiral sources correspond to neutron star binaries, with $m_{NS}\sim 1.4 m_{\odot}$, ($m_{\odot}$ is the solar mass) or perhaps black holes with typical mass $m_{BH}\sim 10 m_{\odot}$.    Since the time evolution is non-relativistic, it is to a good approximation treated in terms of Newtonian gravity.   In particular, the orbital parameters are related by 
\begin{equation}
v^2\sim {G_N m\over r}\equiv {r_s\over 2 r},
\end{equation}   
where $r$ is the orbital radius, and we have introduced the Schwarzschild radius $r_s=2 G_N m$.   The frequency of the radiation emitted during this phase is of order the orbital frequency $2\pi\nu\sim v/r$, so for a binary inspiral that scans the LIGO band, the typical orbital radius is measured in kilometers
\begin{eqnarray}
\label{eq:rad}
r(10\mbox{ Hz})\sim 300\mbox{ km} \left({m\over m_\odot}\right)^{1/3} &\rightarrow& r(1\mbox{ kHz})\sim 14\mbox{ km} \left({m\over m_\odot}\right)^{1/3},
\end{eqnarray}
which is therefore well separated from the size of the compact objects, $r_s\sim 1\mbox{ km}\,  m/m_\odot$.  The expansion parameter $v$ also evolves as the binary sweeps the detector frequency band:
\begin{eqnarray}
\label{eq:v}
v(10\mbox{ Hz})\sim 0.06 \left({m\over m_\odot}\right)^{1/3} &\rightarrow& v(1\mbox{ kHz})\sim 0.3 \left({m\over m_\odot}\right)^{1/3},
\end{eqnarray}
indicating that the velocity is a good expansion parameter up to the last few cycles of orbit seen by the detector.    As $v\rightarrow 1$, the inspiral phase ends, and the evolution must be treated by numerical simulations.

During the inspiral phase, the bound orbit is unstable to the emission of gravitational radiation.    The power emitted in gravitational waves is well approximated by the quadrupole radiation formula.    Taking the orbit to be circular, the power radiated is
\begin{equation}
{dE\over dt} = {32\over 5} G^{-1}_N v^{10}.
\end{equation}
The mechanical energy of the binary is just  $E=-{1\over 2} mv^2$  to leading order in $v$, so by conservation of energy
\begin{equation}
{d\over dt}\left(-{1\over 2} m v^2\right) = -{32\over 5} G^{-1}_N v^{10},
\end{equation}
we obtain an estimate for the duration of the inspiral event seen by LIGO
\begin{equation}
\label{eq:t}
\Delta t = {5\over 512} r_s\left[{1\over v^8_i}-{1\over v^8_f}\right]\sim 5\mbox{ min.}\left({m\over m_\odot}\right)^{-8/3},
\end{equation}
and the number of orbital cycles as the signal sweeps the detector band
\begin{equation}
\label{eq:cycle}
N\sim \int^{t_f}_{t_i} \omega(t) dt = {1\over 32}\left[{1\over v^5_i}-{1\over v^5_f}\right]\sim 4\times 10^4 \left({m\over m_\odot}\right)^{-5/3}\mbox{ radians},
\end{equation}
with $\omega$ the orbital angular frequency.

Eqs.~(\ref{eq:rad}), (\ref{eq:v}), (\ref{eq:t})   give an estimate for the typical length, velocity, and duration of an inspiral event that sweeps the LIGO frequency band.       Inspiral events seen by LISA follow the same dynamics.    However, because LISA will operate in a frequency range complementary to LIGO, $10^{-5}\mbox{ Hz}<\nu<1\mbox{ Hz}$,  the sources correspond to objects of much large mass (e.g., BH/BH binaries with $m_{BH}\sim 10^{5-8} m_{\odot}$ or BH/NS systems with $m_{BH}\sim 10^{5-7}$).

For either LIGO or LISA, the number of orbital cycles spent in the detector frequency band is quite large, see e.g. Eq.~(\ref{eq:cycle}).    Thus even a slight deviation between theoretical calculations of the gravitational wave phase (the waveform ``templates'') and the data will become amplified over the large number of cycles of evolution.   Consequently, inspiral wave signals carry detailed information about gravitational dynamics.    In fact, it has been determined that LIGO will be sensitive to corrections that are order $v^6$ in the velocity expansion beyond the leading order quadrupole radiation predictions~\cite{kip}.   

The procedure for computing corrections to the motion of the binary system in the non-relativistic limit by an iterated expansion of the Einstein equations is called the post-Newtonian expansion of general relativity~\cite{PN}.     What makes the calculations difficult (and interesting) is that there is a proliferation of physically relevant effects occurring at \emph{different length scales}, 
\begin{eqnarray}
\nonumber
r_s &=&\mbox{size of compact objects},\\
\nonumber
r &=&\mbox{orbital radius},\\
\nonumber
\lambda &=& \mbox{wavelength of emitted radiation},
\end{eqnarray}
all controlled by the same expansion parameter $v$:
\begin{eqnarray}
{r_s\over r} \sim v^2 & {r\over\lambda}\sim v,
\end{eqnarray}
where the first estimate follows from Kepler's law and the second from the multipole expansion of the radiation field coupled to non-relativistic sources.   Thus to do systematic calculations at high orders in $v$, it is necessary to deal with physics at many different scales.   A convenient way of doing this is to construct a chain of EFTs that capture the relevant physics at each scale separately.

The goal of these lectures is to recast the problem of computing gravitational wave observables for non-relativistic sources  in terms of EFTs.   In the next section we briefly review the EFT logic, using a standard particle physics example, the low energy dynamics of Goldstone bosons in a theory with spontaneously broken global symmetry, to illustrate the main ideas.    In sec.~\ref{sec:grEFT}, we turn to the EFT formulation of the binary inspiral problem.    The presentation follows~\cite{GnR1}, which is based on similar EFTs for non-relativistic bound states in QED and QCD~\cite{NRQCD}.  However, the focus here is not on detailed calculations, but in describing how to ``integrate out'' physics at each of the scales $r_s$, and $r$ to obtain a theory with well defined rules for calculating gravitational wave observables (at the scale $\lambda$) as an expansion in $v$.    Possible extensions of the ideas presented here to other problems in gravitational wave physics are discussed in the conclusions, sec.~\ref{sec:conc}.

\subsection{Effective field theories:   a review}

EFTs are indispensable for treating problems that simultaneously involve two or more widely separated scales.   As a typical application, suppose one is interested in calculating the effects of some sort of short distance physics, characterized by a scale $\Lambda$, on the dynamics at a low energy scale $\omega\ll \Lambda$.   In the EFT description of such a problem, the effects of $\Lambda$ on the low energy physics become simple,   making it possible to construct a systematic expansion in the ratio $\omega/\Lambda\ll 1$.   Here we briefly review the basic ideas that go into the construction and applications of EFTs.   More detailed reviews can be found in refs.~\cite{polchinski,kaplan,aneesh,ira}.

The key insight that makes EFTs possible is the following:    consider for instance a field theory with light degrees of freedom collectively denoted by $\phi$ (e.g, a set of massless fields) and some heavy fields $\Phi$ with mass near the UV scale $\Lambda$.   The interactions of these fields are described by some action functional $S[\phi,\Phi]$.   Suppose also that we are only interested in describing experimental observables involving only the light modes $\phi$.   Then it makes sense to integrate out the modes $\Phi$ to obtain an effective action that describes the interactions of the light fields among each other
\begin{equation}
\label{eq:WilsonPI}
e^{i S_{\mathit{eff}}[\phi]} = \int D\Phi(x) \, e^{i S[\phi,\Phi]}.
\end{equation}   
It turns out that in general, $S_{\mathit{eff}}[\phi]$ can be expressed as a \emph{local} functional of the fields $\phi(x)$, 
\begin{equation}
\label{eq:local}
S_{\mathit{eff}}[\phi]= \sum_i c_i  \int  d^4 x \,{\cal O}_i(x),
\end{equation}
for (in general an infinite number) local operators ${\cal O}_i(x)$.    The coefficients $c_i$ are usually referred to as ``Wilson coefficients''.  If  ${\cal O}_i(x)$ has mass dimension $\Delta_i$, then  the Wilson coefficients, evaluated at a renormalization point $\mu$ of order $\Lambda$,  scale as powers of $\Lambda$
\begin{equation}
\label{eq:scale}
c_i(\mu=\Lambda)= {\alpha_i\over \Lambda^{\Delta_i-4}},
\end{equation}  
with $\alpha_i\sim{\cal O}(1)$.   From this observation we conclude that the short distance physics can have two types of effects on the dynamics at energies $\omega\ll \Lambda$:
\begin{itemize}
\item Renormalization of the coefficients of operators with mass dimension $\Delta\leq 4$.
\item Generation of an infinite tower of irrelevant (i.e. $\Delta>4$) operators with coefficients scaling as in Eq.~(\ref{eq:scale}).
\end{itemize}

This result regarding the structure of the low energy dynamics encoded in $S_{\mathit{eff}}[\phi]$ is usually referred to as \emph{decoupling}.   This concept has its roots in the work of K. Wilson on the renormalization group~\cite{wilson}.   Decoupling as used by practicing field theorists was first made rigorous in~\cite{ac}.   The idea is useful because it states that the dependence of the low energy physics on the scale $\Lambda$ is extremely simple.   All UV dependence appears directly in the coefficients of the effective Lagrangian, and determining the $\Lambda$ dependence of an observable follows from \emph{power counting} (essentially a generalized form of dimensional analysis).   Effective Lagrangians are typically used in one of two ways:

\begin{enumerate}
\item The ``full theory'' $S[\phi,\Phi]$ is known:   In this case, integrating out the heavy modes gives a simple way of systematically analyzing the effects of the heavy physics on low energy observables.  Because only the low energy scale appears explicitly in the Feynman diagrams of the EFT, amplitudes are easier to calculate and to power count than in the full theory.   

{\bf Examples:}    Integrating out the $W,Z$ bosons from the $SU(2)_L\times U(1)_Y$ Electroweak Lagrangian at energies $E\ll m_{W,Z}$ results in the Fermi theory of weak decays plus corrections suppressed by powers of $E^2/m^2_{W,Z}$.   It is easier to analyze electromagnetic or QCD corrections to weak decays in the four-Fermi theory than in the full Electroweak Lagrangian, as the graphs of Fig.~\ref{fig:FF} indicate.   See, e.g., ref.~\cite{weak}.       Another example is the use of EFTs to calculate  heavy particle threshold corrections to low energy gauge couplings~\cite{thresh} .   This has applications, for instance, in Grand Unified Theories and in QCD.

\begin{figure}
\def\size{5cm}
\hbox{\vbox{\hbox to \size {\hfil \includegraphics[width=4cm]{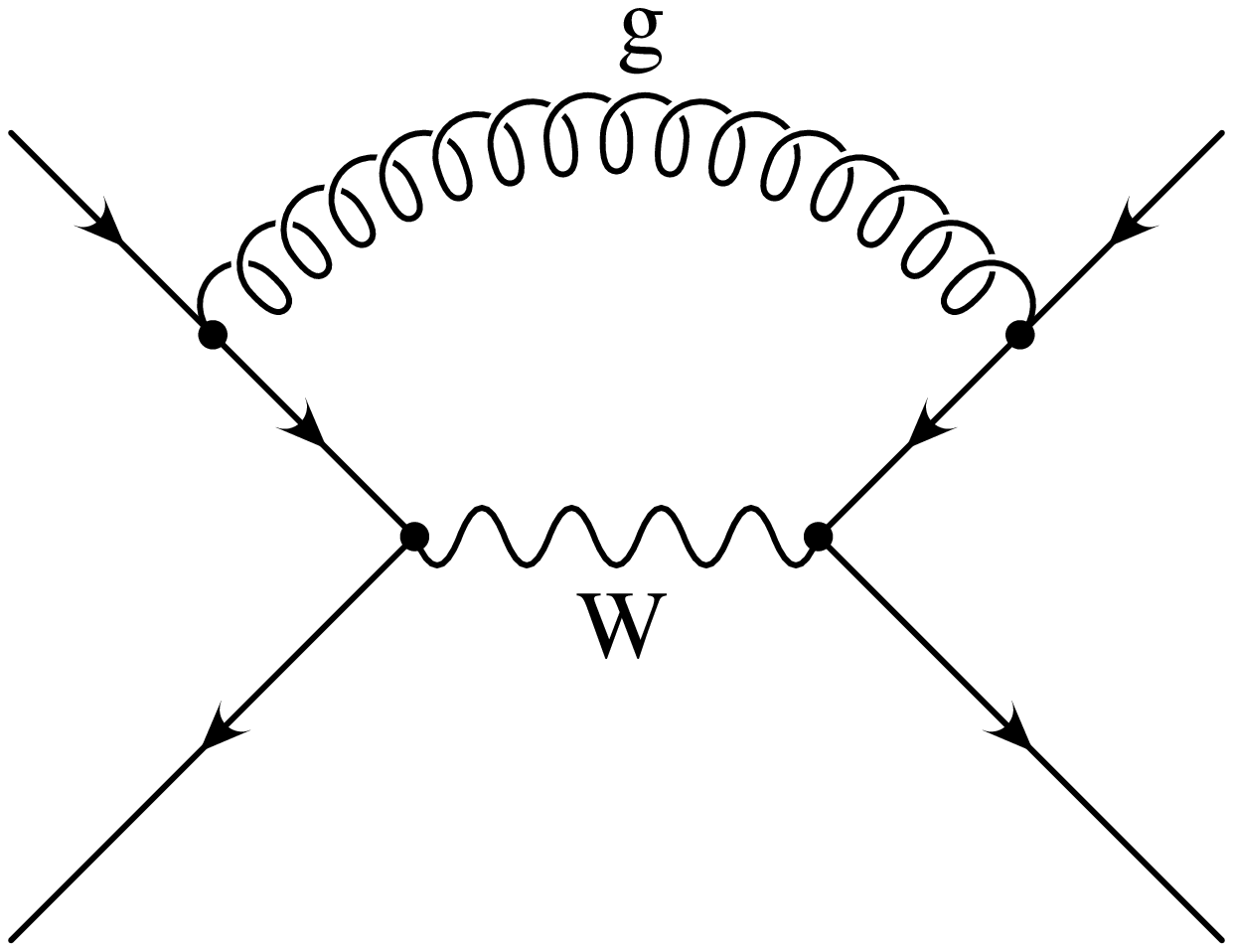} \hfil }\hbox to \size {\hfil(a)\hfil}}
\vbox{\hbox to \size {\hfil \includegraphics[width=2.7cm]{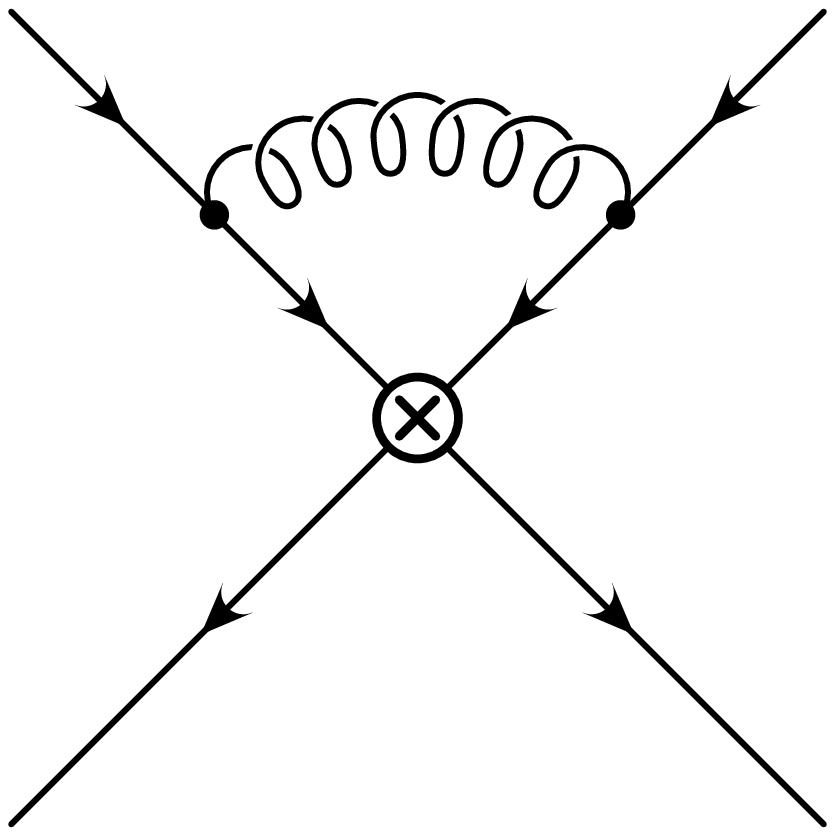} \hfil}\hbox to \size {\hfil(b)\hfil}}}
\caption{QCD correction to a typical four-Fermi process in the Standard Model, calculated in (a) the full theory with propagating $W$ bosons, and (b) the effective Fermi theory of weak interactions.    Graph (b) reproduces graph (a) up to corrections suppressed by powers of $E^2/m^2_W\ll 1$. \label{fig:FF}}
\end{figure}

 \item The full theory is unknown (or known but strongly coupled):    Whatever the physics at the scale $\Lambda$ is, by decoupling it must manifest itself at low energies as an effective Lagrangian of the form Eq.~(\ref{eq:local}).   If the symmetries (eg. Poincare, gauge, global) that survive at low energies are known, then the operators ${\cal O}_i(x)$ appearing in $S_{\mathit eff}[\phi]$ must respect those symmetries.   Thus by writing down an effective Lagrangian containing the most general set of operators consistent with the symmetries, we are necessarily accounting for the UV physics in a completely \emph{model independent way}.    

{\bf Examples}:   The QCD chiral Lagrangian below the scale $\Lambda_{\chi SB}$ of $SU(3)_L\times SU(3)_R\rightarrow SU(3)_V$ chiral symmetry breaking~\cite{chi1,chi2}.   Here the full theory, QCD, is known, but because $\Lambda_{\chi SB}$ is of order the scale $\Lambda_{QCD}\sim 1\, \mbox{GeV} $ where the QCD coupling is strong, it is impossible to perform the functional integral in Eq.~(\ref{eq:PI}) analytically.   Another example is general relativity below the scale $m_{Pl}\sim 10^{19}\, \mbox{GeV}$.    This theory can be used to calculate, e.g., graviton-graviton scattering at energies $E\ll m_{Pl}$.  Above those energies, however, scattering amplitudes calculated in general relativity start violating unitarity bounds, and the effective field theory necessarily breaks down.   Thus general relativity is an effective Lagrangian for quantum gravity below the strong coupling scale $m_{Pl}$.  The EFT interpretation of general relativity is reviewed in more detail in refs.~\cite{donoghue,burgess}   

Finally, it is believed that the Standard Model itself is an effective field theory below scales of order $\Lambda=1\,\mbox{TeV}$ or so (see lectures by H. Murayama at this school).   This scale manifests itself indirectly, in the form of $SU(2)_L\times U(1)_Y$ gauge invariant operators of dimension $\Delta>4$ constructed from Standard Model fields~\cite{EWEFT}, certain linear combinations of which have been constrained experimentally using collider data from the LEP experiments at CERN and from SLD at SLAC (see~\cite{skiba} for a recent analysis of precision electroweak constraints using effective Lagrangians).  If there is indeed new physics at the TeV scale, it will be seen directly, at the CERN LHC which is due to come on line in the next few years.
\end{enumerate}

In either of these two classes of examples, integrating out the heavy physics as in Eq.~(\ref{eq:PI}) results in a an effective Lagrangian that contains, in general, an \emph{infinite} number of operators ${\cal O}_i(x)$.   However, because
\begin{enumerate}
\item An operator ${\cal O}(x)$ with $[{\cal O}]=\Delta_{\cal O}>4$ contributes to an observable at relative order 
\begin{equation}
\left({\omega\over\Lambda}\right)^{\Delta_{\cal O}-4}\ll 1,
\end{equation}
\item A given observable can only be determined up to a finite experimental resolution $\epsilon\ll 1$,
\end{enumerate}
one may typically truncate the series in Eq.~(\ref{eq:local}) after a finite number of operators, those with mass dimension $\Delta\leq N+4,$ where
\begin{equation}
\epsilon=\mbox{expt. error} = \left({\omega\over\Lambda}\right)^N.
\end{equation}
Therefore $S_{eff}[\phi]$ is predictive as long as there are more observables than operators with mass dimension $\Delta\leq N+4$.

To make these general remarks more concrete, it is worth studying a toy example in some detail.   We'll consider the quantum field theory of a single complex scalar field $\phi(x)$ with Lagrangian (the ``full theory'')
\begin{equation}
\label{eq:fullU(1)}
{\cal L} = |\partial_\mu \phi|^2 - V(\phi),
\end{equation}
where
\begin{equation}
V(\phi)={\lambda\over 2}\left(|\phi|^2 - {v^2/2}\right)^2.
\end{equation}
This theory is invariant under the global $U(1)$ symmetry
\begin{equation}
U(1) : \phi(x)\rightarrow e^{i\alpha} \phi(x),
\end{equation}
for some constant phase $\alpha$, as well as a discrete charge conjugation symmetry
\begin{equation}
C: \phi(x)\rightarrow \phi^{*}(x).
\end{equation}
Classically, the ground state of this theory is determined by the minimum of $V(\phi)$.  Because of the $U(1)$ symmetry the ground state is degenerate, and the vacuum manifold is the circle $|\phi|^2 = v^2/2$ in the complex $\phi$ plane.   To study fluctuations about the vacuum, we expand the fields about any of these (equivalent) vacuua, for instance the point
\begin{equation}
\langle \phi\rangle = {v\over\sqrt{2}}.
\end{equation} 
Expanding about this point spontaneously breaks the $U(1)$ symmetry, resulting in one Goldstone boson.    It is convenient to write the original field $\phi(x)$ as
\begin{equation}
\phi(x)={1\over\sqrt{2}}\left(v+\rho(x)\right) e^{i\pi(x)/v}.
\end{equation}
Under the original symmetries, the new fields transform as 
\begin{equation}
U(1): 
\left\{\begin{array}{l}
 \rho(x)\rightarrow\rho(x),\\
  \pi(x)/v\rightarrow \pi(x)/v+\alpha,
\end{array}\right.  
\end{equation}
 and 
 \begin{equation}
 C: \left\{\begin{array}{l}
 \rho(x)\rightarrow\rho(x),\\
  \pi(x)\rightarrow-\pi(x).
  \end{array}\right.
 \end{equation} 
The Lagrangian in terms of the new fields is given by 
\begin{equation}
{\cal L} = {1\over 2}(\partial_\mu \rho)^2 + {1\over 2} \left(1+{\rho\over v}\right)^2 (\partial_\mu \pi)^2 -{\lambda\over 2} (v\rho + \rho^2/2)^2,
\end{equation}
so the spectrum of excitations about the vacuum consists of a ``modulus'' field $\rho(x)$ with tree level mass given by $m^2_\rho=\lambda v^2$ and the Goldstone boson $\pi(x)$ with $m_\pi=0$.  (We take $\lambda\ll 1$ so that a perturbative treatment is valid).

Suppose that we are interested in working out the predictions  of this theory at energies $\omega \ll m_\rho$.  At such scales, only the dynamics of the massless field $\pi(x)$ is non-trivial, and following our general discussion it is convenient to write an effective Lagrangian.   The general structure of this effective Lagrangian is dictated by the symmetries of the original theory.   In particular the $U(1)$, realized non-linearly as the shift symmetry $\pi(x)/v\rightarrow \pi(x)/v+\alpha$, restricts the effective Lagrangian to be a function of $\partial_\mu \pi$ only.   Therefore the symmetries of the low energy theory alone explain a number of consequences of the low energy dynamics, for instance
\begin{itemize}
\item The field $\pi(x)$ must be massless, since a mass term would break the shift symmetry.   This is just  the statement of Goldstone's theorem in the context of the low energy EFT.
\item The field $\pi(x)$ is derivatively self-coupled.  This implies in particular that scattering amplitudes are ``soft'', vanishing as powers of the typical energy $\omega$ in the limit $\omega\rightarrow 0$.
\end{itemize}
In addition to the constraints from the non-linearly realized $U(1)$, there is also a constraint from the charge conjugation symmetry $\pi\rightarrow -\pi$ which says that the EFT must be even in $\pi$.    Thus the effective Lagrangian must be of the general form
\begin{equation}
\label{eq:loeft}
{\cal L}_{EFT} = {1\over 2}(\partial_\mu\pi)^2 + {c_8\over 4\Lambda^4} (\partial_\mu\pi\partial^\mu\pi)^2 + \cdots,
\end{equation}
where only the leading operators with two and four $\pi$'s have been displayed.   In this equation $\Lambda\sim m_\rho$ and $c_8$ is some dimensionless constant of order $\lambda$.

\subsubsection{Matching}
Since we know the full theory, Eq.~(\ref{eq:fullU(1)}), it is possible to explicitly perform the functional integral of Eq.~(\ref{eq:WilsonPI}) in order to obtain the EFT parameters like $\Lambda$ and $c_8$ in terms of the couplings $\lambda,v$ appearing in the full theory Lagrangian.    Rather than calculating this integral, the low energy parameters are fixed in practice through a procedure referred to as \emph{matching}.  

To perform a matching calculation, one simply calculates some observable, for instance a scattering amplitude among the light particles, in two ways.     First one calculates the amplitude in the full theory, expanding the result in powers of $\omega/\Lambda$.  One then calculates the same quantity in the effective field theory, adjusting the EFT parameters in order to reproduce the full theory result.  

\begin{figure*}
\def\size{5cm}
\includegraphics[width=5cm]{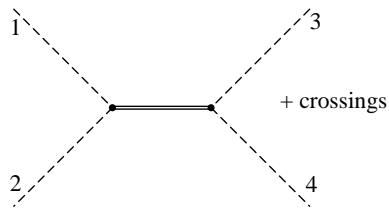} 
\caption{Leading order contribution to $\pi\pi\rightarrow\pi\pi$ scattering in the full theory.   The intermediate double line corresponds to the $\rho$ field propagator. \label{fig:pptopp}}
\end{figure*} 

As an example, consider $\pi\pi\rightarrow \pi\pi$ scattering in our toy model.    In the full theory, the amplitude to leading order in $\lambda$  is given by the diagrams of Fig.~\ref{fig:pptopp}.   Reading off the Feynman rules from the full theory Lagrangian, Eq.~(\ref{eq:fullU(1)}), one finds
\begin{eqnarray}
i{\cal A}_{\mathit full} = \left[{2 i\over v} k_1\cdot k_2\right] {i\over s-m^2_\rho} \left[ {2 i\over v} k_3\cdot k_4\right]  + \mbox{crossings},
\end{eqnarray}
where $k_1, k_2$ are the initial (incoming) momenta, and $k_3, k_4$ are the final (outgoing) momenta.  Introducing the usual Mandelstam variables $s=(k_1+k_2)^2$, $t=(k_1-k_3)^2$, $u=(k_1-k_4)^2$, and expanding in $s,t,u\ll m^2_\rho,$ this becomes
\begin{equation}
i{\cal A}_{\mathit full}\simeq {4 i\over v^2 m_\rho^2} \left[\left({s\over 2}\right)^2 + \left({t\over 2}\right)^2 + \left({u\over 2}\right)^2\right] + {\cal O}(s^3/m^6_\rho)
\end{equation}
 In the effective field theory, the amplitude arises from the leading dimension eight interaction in Eq.~(\ref{eq:loeft}), 
\begin{equation}
i{\cal A}_{EFT} = {i c_8\over \Lambda^4} \left[\left({s\over 2}\right)^2 + \left({t\over 2}\right)^2 + \left({u\over 2}\right)^2\right].
\end{equation}
Thus taking $\Lambda=m_\rho$, the matching condition for the coefficient $c_8$ is given by
\begin{equation}
c_8(\mu=\Lambda) = 4\lambda + {\cal O}(\lambda^2).
\end{equation}

\subsubsection{Corrections to the matching coefficients}

In general, the coefficients in the effective Lagrangian are calculable as a series expansion in the parameters of the full theory.   In our example, the coefficients receive corrections at all orders in the parameter $\lambda\ll 1$, of which the tree level matching calculation in the above example gives only the first term.   It is instructive to see how one goes about computing corrections to the Wilson coefficients, as this illustrates certain general properties of effective field theories.

\begin{figure}
\def\size{5cm}
\hbox{\vbox{\hbox to \size {\hfil \includegraphics[width=4cm]{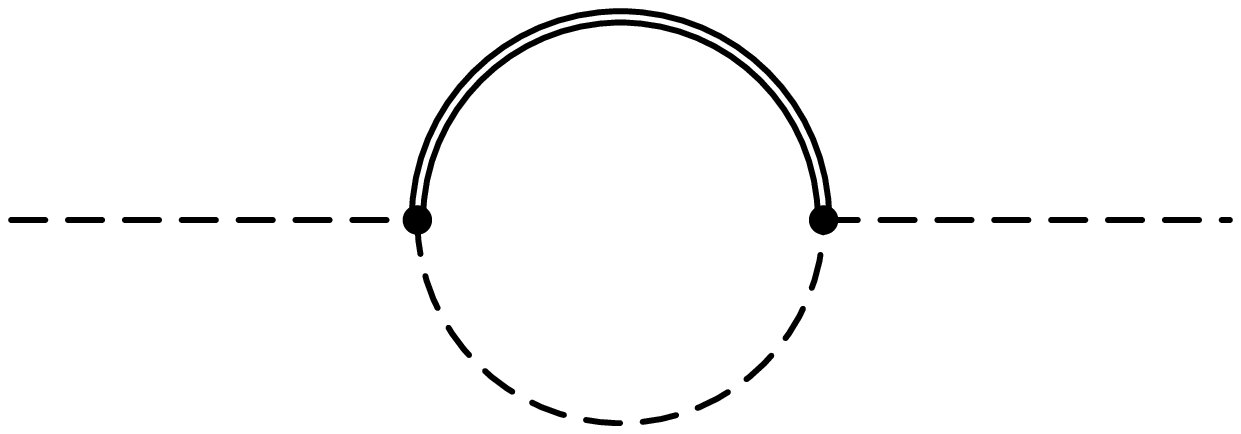} \hfil }\hbox to \size {\hfil(a)\hfil}}
\vbox{\hbox to \size {\hfil \includegraphics[width=4cm]{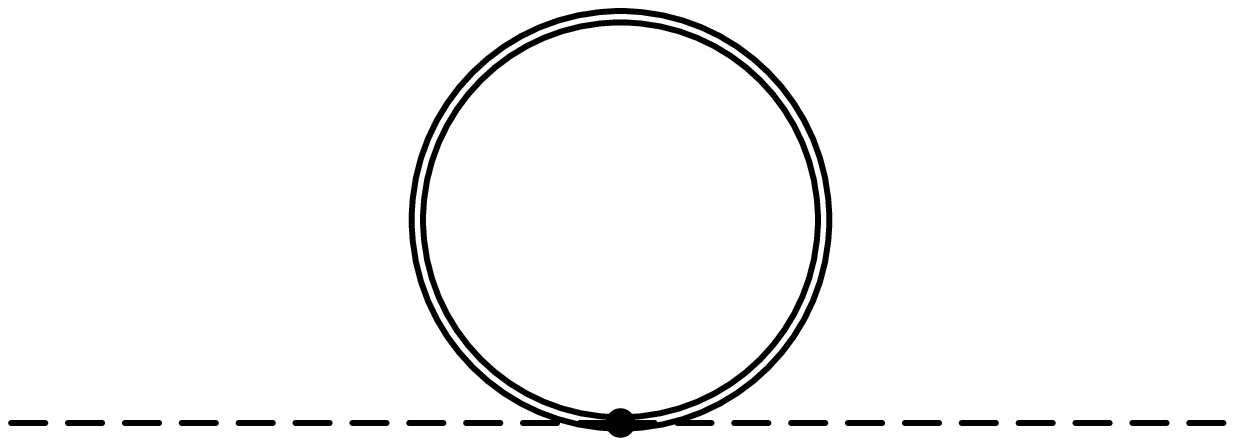} \hfil}\hbox to \size {\hfil(b)\hfil}}}
\vspace{0.25cm}
\hbox{\vbox{\hbox to \size {\hfil \includegraphics[width=4cm]{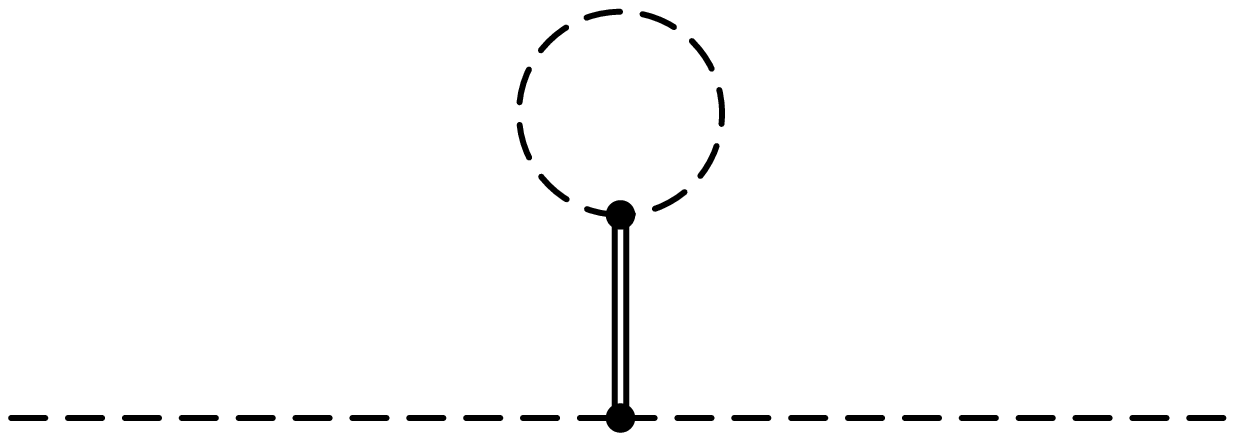} \hfil }\hbox to \size {\hfil(c)\hfil}}
\vbox{\hbox to \size {\hfil \includegraphics[width=4cm]{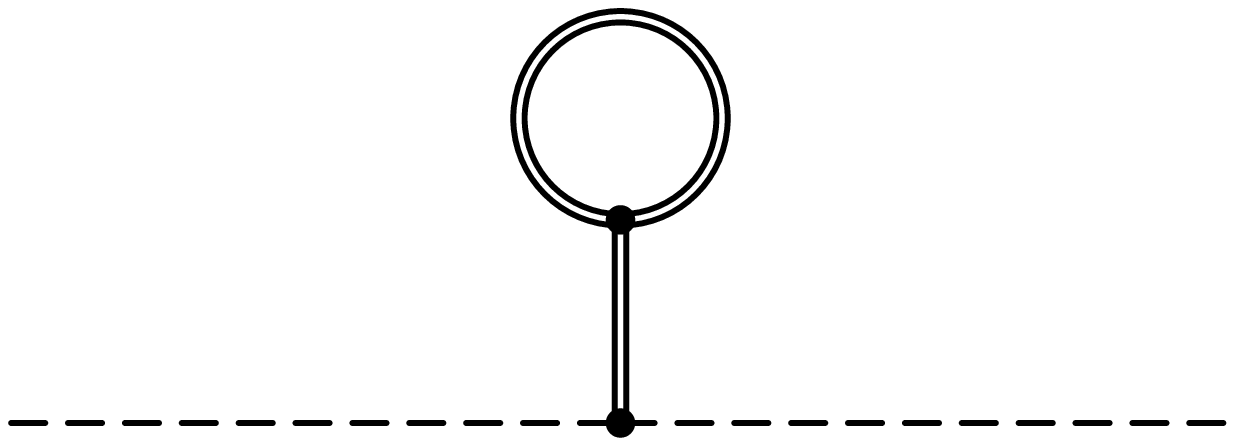} \hfil}\hbox to \size {\hfil(d)\hfil}}}
\caption{Feynman graphs contributing to the $\pi$ field self-energy in the full theory with a propagating $\rho$ field.  \label{fig:2ptfull}}
\end{figure}

In the example above, consider for instance the coefficient of the kinetic term for the $\pi$ field in the low energy EFT,
\begin{equation}
{\cal L}_{EFT}={1\over 2} Z (\partial_\mu \pi)^2 + \cdots.
\end{equation} 
To leading order $Z=1$.    Corrections of order $\lambda$ are obtained by comparing the one-loop $\pi$ field propagator in the full and the effective theories, adjusting $Z$ so that the two calculations agree.   Rather than calculate the full propagator, it is enough to consider the ``one light-particle irreducible''  two-point function.    In the full theory this is the sum of graphs that cannot be made disconnected by cutting one $\pi$ field propagator.   In the EFT, it is just the usual 1PI graphs.  The relevant diagrams are shown in the full and effective theories in Fig.~\ref{fig:2ptfull}, and Fig.~\ref{fig:2pteft} respectively.   In the full theory we find at one-loop
\begin{eqnarray}
\nonumber
\mbox{Fig.~\ref{fig:2ptfull}(a)} &=&  {i\lambda\over 16\pi^2} k^2 \left[{1\over\epsilon} -\gamma+{3\over 2} -\ln \left({m^2_\rho\over 4\pi\mu^2}\right)\right] + {\cal O}(k^4/m^2_\rho),\\
\end{eqnarray}
\begin{eqnarray}
\mbox{Fig.~\ref{fig:2ptfull}(b)} &=& - {i\lambda\over 16\pi^2} k^2 \left[{1\over\epsilon} -\gamma+1 -\ln \left({m^2_\rho\over 4\pi\mu^2}\right)\right],
\end{eqnarray}
where the graphs have been regulated by dimensional regularization in $d=4-2\epsilon$ dimensions.  
The third diagram is 
\begin{equation}
\mbox{Fig.~\ref{fig:2ptfull}(c)} = {2 i \lambda k^2 \over m^4_\rho} \int {d^d q\over (2\pi)^d} =0,
\end{equation} 
where we have made use of the fact that in dimensional regularization
\begin{equation}
\label{eq:dr}
\lim_{n\rightarrow 0} \int {d^d q\over (2\pi)^d}  {1\over (q^2)^n}  = \lim_{n\rightarrow 0} {i\over (4\pi)^{d/2}} {\Gamma(n-d/2)\over  \Gamma(n)} (0)^{d/2-n}\rightarrow 0.
\end{equation}
The last diagram is given by 
\begin{equation}
\mbox{Fig.~\ref{fig:2ptfull}(d)}= {3 i\lambda\over 16\pi^2} k^2\left[{1\over\epsilon} -\gamma +1-\ln \left({m^2_\rho\over 4\pi\mu^2}\right)\right].
\end{equation}
Note that the $1/\epsilon$ divergences in Fig.~\ref{fig:2ptfull}(a) and Fig.~\ref{fig:2ptfull}(b) cancel each other.   This is equivalent to the statement that the full theory does not have wavefunction renormalization at one-loop order (it is just  $\phi^4$ after all).    On the other hand, the tadpole graph in Fig.~\ref{fig:2ptfull}(d), which corresponds to a radiative correction to the VEV of the original field $\phi$ requires a tree graph (not shown in the figure) with a counterterm insertion to regulate the logarithmic divergence.    After subtracting this $1/\epsilon$ pole using the counterterm contribution, one finds the following result for the renormalized $\pi$ field self-energy in the full theory at one-loop (working in the $\overline{MS}$ scheme):
\begin{equation}
-i\Pi_{\pi}(k^2) = {i\lambda\over 16\pi^2} k^2 \left[{7\over 2} -3 \ln\left({m^2_\rho\over \mu^2}\right)\right]+{\cal O}(k^4/m^2_\rho).
\end{equation}

In the EFT, the self-energy of the Goldstone boson is given by the graphs of Fig.~\ref{fig:2pteft}.   These are:
\begin{figure}
\def\size{5cm}
\hbox{\vbox{\hbox to \size {\hfil \includegraphics[width=4cm]{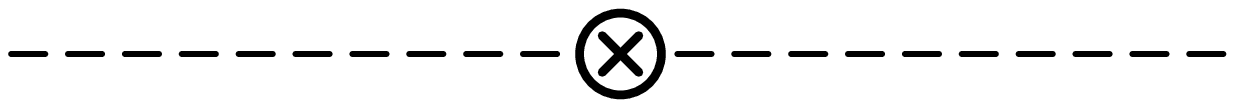} \hfil }\hbox to \size {\hfil(a)\hfil}}
\vbox{\hbox to \size {\hfil \includegraphics[width=4cm]{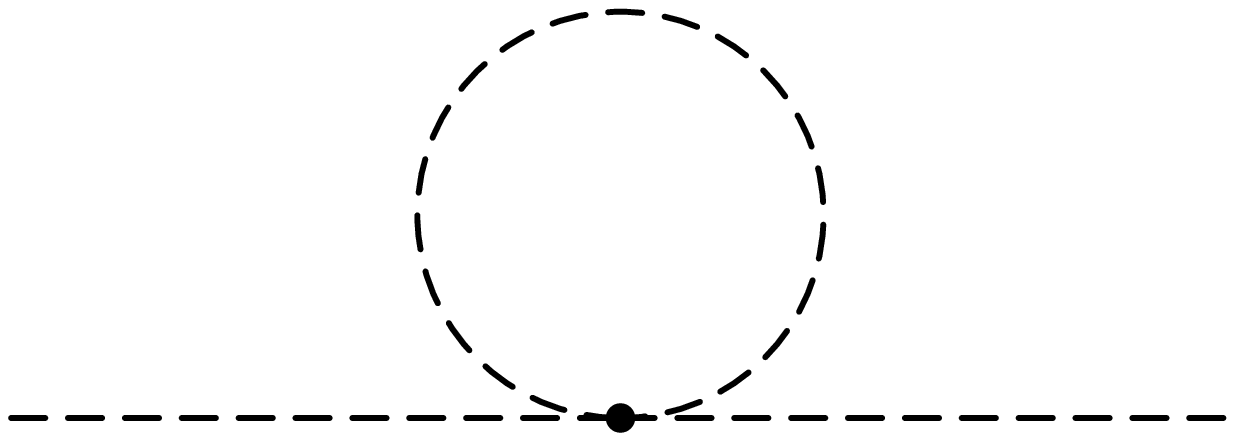} \hfil}\hbox to \size {\hfil(b)\hfil}}}
\caption{Feynman graphs contributing to the $\pi$ field self-energy in the low energy EFT.   Graph (b), including one insertion of the coupling $c_8$ vanishes in dimensional regularization.  \label{fig:2pteft}}
\end{figure}
\begin{equation}
\mbox{Fig.~\ref{fig:2pteft}(a)}  = i(Z-1) k^2,
\end{equation}
and
\begin{equation}
\mbox{Fig.~\ref{fig:2pteft}(b)} =0,
\end{equation}
by Eq.~(\ref{eq:dr}).  Setting the full and EFT results equal, we find
\begin{equation}
Z=1 + {\lambda\over 16\pi^2}\left[ {7\over 2} -3 \ln\left({m^2_\rho\over \mu^2}\right)\right] +{\cal O}(\lambda^2).
\end{equation}

The one-loop matching calculation presented here serves to illustrate two general features of matching calculations:
\begin{itemize}
\item Integrating out heavy modes not only generates new irrelevant (dimension $\Delta>4$) operators in the effective theory.   It also renormalizes the coefficients of the operators with $\Delta\leq 4$.
\item Loop graphs in matching calculations typically contain logarithms of $m_\rho/\mu$.    In order to avoid possible large logarithms that could render perturbation theory invalid, one must choose a matching scale $\mu$ that is of the same order as the masses of the fields that are being integrated out.   
\end{itemize}
This last point in particular implies that it is the Wilson coefficients $c_i(\mu\sim\Lambda)$ that exhibit simple scaling in powers of the UV scale $\Lambda$.   Given the RG equations for the full theory coupling constants,  one may also use any other $\mu > \Lambda$  as the matching scale.    Possible large logarithms that may arise from this choice are then resummed by RG running.  For example, using
\begin{equation}
\mu {dZ(\mu)\over d\mu} = {3\lambda(\mu) \over 8\pi^2} + {\cal O}(\lambda^2),
\end{equation}  
and the one-loop RG equation for the full theory coupling constant $\lambda$,
\begin{equation}
\mu {d  \lambda(\mu) \over d\mu}= {5\lambda^2(\mu)\over 8\pi^2},
\end{equation}
one finds the relation
\begin{equation}
Z(\mu) = Z(\Lambda) +{3\over 5} \ln {\lambda(\mu)\over\lambda(\Lambda)}.
\end{equation}
Thus in this theory, using the ``wrong'' matching scale has a relatively mild effect, although this is not generally true in other theories, e.g. QCD.   In any case, it is simplest in practice to just match directly at $\mu\sim\Lambda$.  

Likewise, to calculate observables at low energy scale $\omega\ll \Lambda$, it is better to evaluate loop graphs in the EFT at a renormalization point $\mu\simeq \omega$, indicating that in the EFT one should use the couplings $c_i(\mu\sim\omega)$.    These can be obtained in terms of the coefficients $c_i(\mu\sim\Lambda)$ obtained through matching by renormalization group (RG) evolution between the scales $\Lambda$ and $\omega$ within the EFT.

\begin{figure}
\includegraphics[width=5cm]{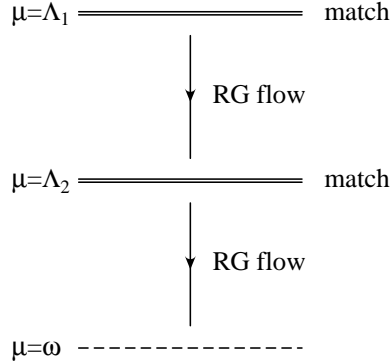}
\caption{Construction of a low energy EFT for a theory with scales $\Lambda_1\gg \Lambda_2\gg\omega$.\label{fig:EFTschem}}
\end{figure}

For a theory with multiple scales, the procedure is similar.  A typical example is shown in Fig.~\ref{fig:EFTschem}, which depicts the construction of an EFT at  a low scale $\omega\ll \Lambda_2\ll \Lambda_1$ starting from a theory of light fields $\phi$ coupled to heavy fields $\Phi_1,$ $\Phi_2$ (masses of order $\Lambda_1$, $\Lambda_2$).   One first constructs an $\mbox{EFT}_1$ for $\phi,$ $\Phi_2$ (regarded as approximately massless) by  integrating out the fields $\Phi_1$.   This generates a theory defined by its coupling constants at the renormalization scale $\mu\sim\Lambda_1$.    This $\mbox{EFT}_1$ is then used to RG evolve the couplings down to the threshold $\mu\sim\Lambda_2$, at which point $\Phi_2$ is treated as heavy and removed from the theory.   This finally generates an $\mbox{EFT}_2$ for the light fields $\phi$ which can be used to calculate at the scale $\omega$.   In $\mbox{EFT}_2$, logarithms of $\omega/\Lambda_2$ can be resummed by RG running\footnote{There are exceptions to the construction described in this paragraph, for example NRQCD, the EFT that describes the non-relativistic limit of the strong interactions.   Although this theory contains several widely separated scales, these are correlated implying that RG running must be performed in one stage, rather than in the two-stage picture presented here~\cite{LMR}.   See~\cite{ira} for a pedagogical review.    This will not be an issue in our discussion of the non-relativistic gravity.}.

{\bf Exercise}:   Calculate the one-loop correction to the Wilson coefficient $c_8(\mu)$.    

\subsubsection{Power counting}

In every EFT, there is a rule for determining which operators are needed at a given order in the expansion parameter $\omega/\Lambda\ll1$.   This rule is called \emph{power counting}.   It is slightly unusual, in that a typical power counting scheme assigns the same counting to tree and loop contributions to a given observable.    This is unlike the situation in renormalizable field theories where the perturbative expansion is equivalent to an expansion in powers of $\hbar$ (the loop expansion).

To see how this works, we develop the power counting rules of our toy EFT.   In this case, power counting is simply a matter of keeping track of operator mass dimensions\footnote{This need not always be the case, as we will see in the next lecture.}.  We simply make the obvious assignment
\begin{eqnarray}
k^\mu\sim\omega\Rightarrow\partial_\mu\sim\omega\Rightarrow x^\mu\sim\omega^{-1}.
\end{eqnarray}
Since the kinetic operator is leading in the expansion, it should scale as
\begin{equation}
\int d^4 x (\partial_\mu\pi)^2 \sim \left({\omega\over\Lambda}\right)^0,
\end{equation}
indicating that $\pi(x)\sim\omega$, and therefore the scaling of an operator is simply its mass dimension.

\begin{figure*}
\def\size{5cm}
\hbox{\vbox{\hbox to \size {\hfil \includegraphics[width=2.5cm]{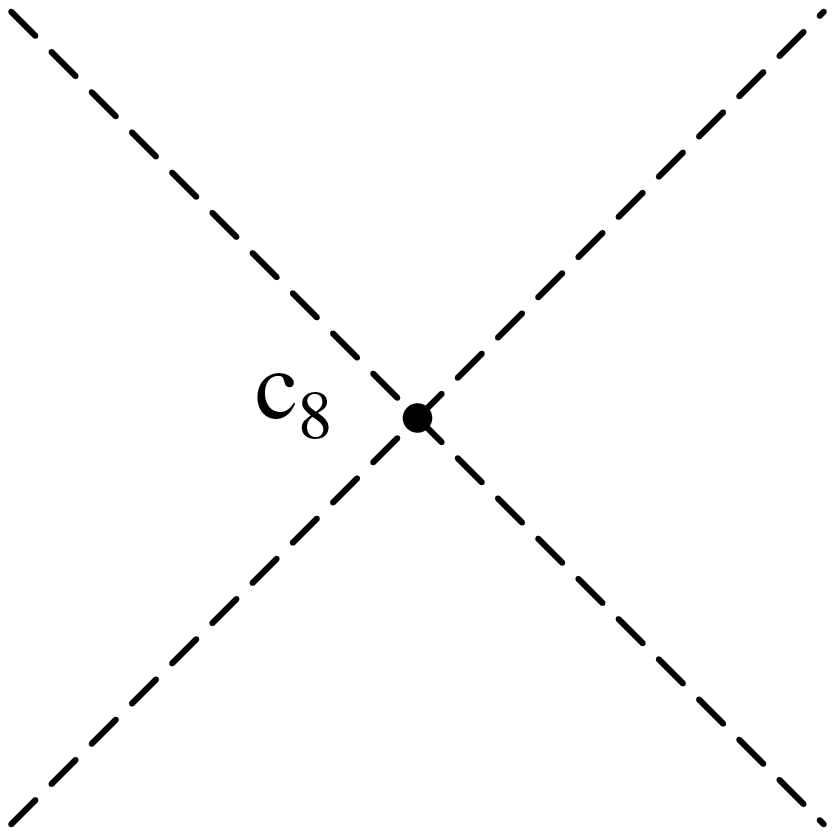} \hfil }\hbox to \size {\hfil(a)\hfil}}
\vbox{\hbox to \size {\hfil \includegraphics[width=2.5cm]{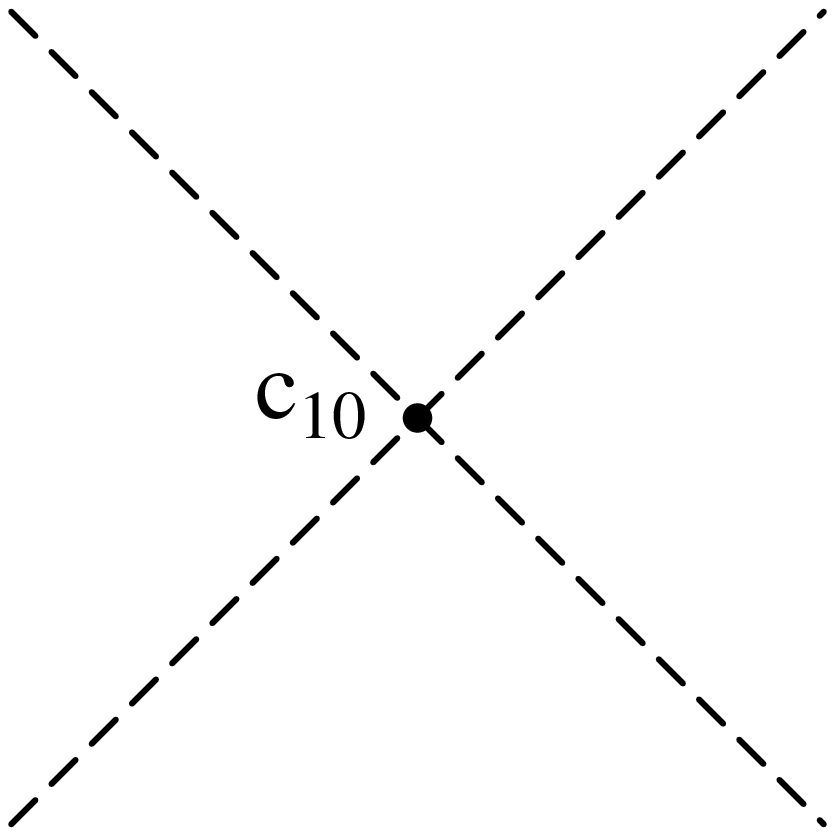} \hfil}\hbox to \size {\hfil(b)\hfil}}}
\hbox{\vbox{\hbox to \size {\hfil \includegraphics[width=2.5cm]{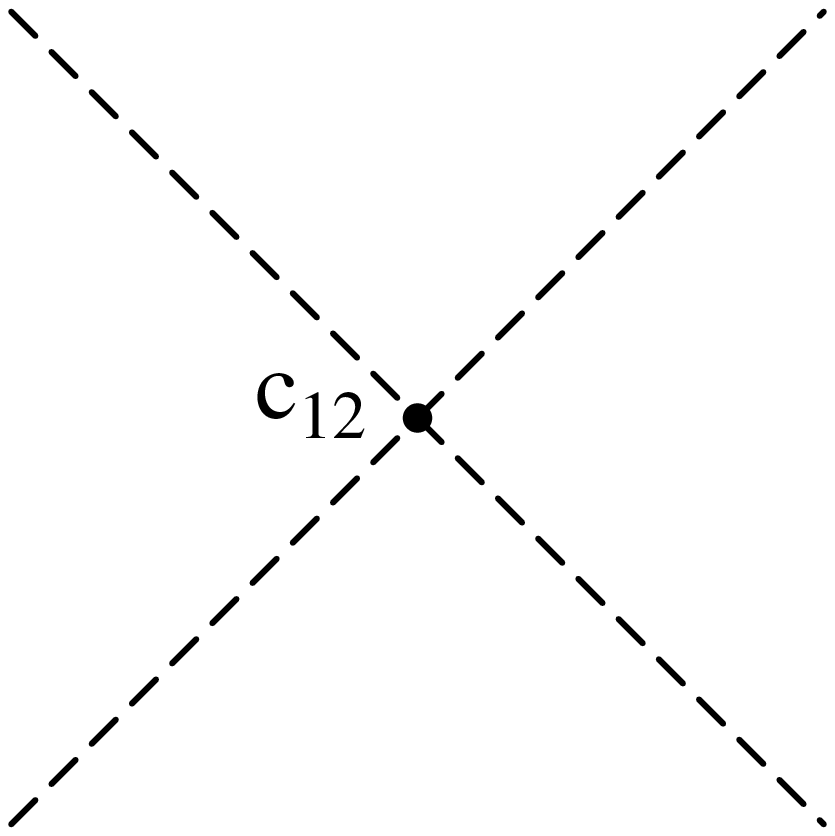} \hfil }\hbox to \size {\hfil(c)\hfil}}
\vbox{\hbox to \size {\hfil \includegraphics[width=5cm]{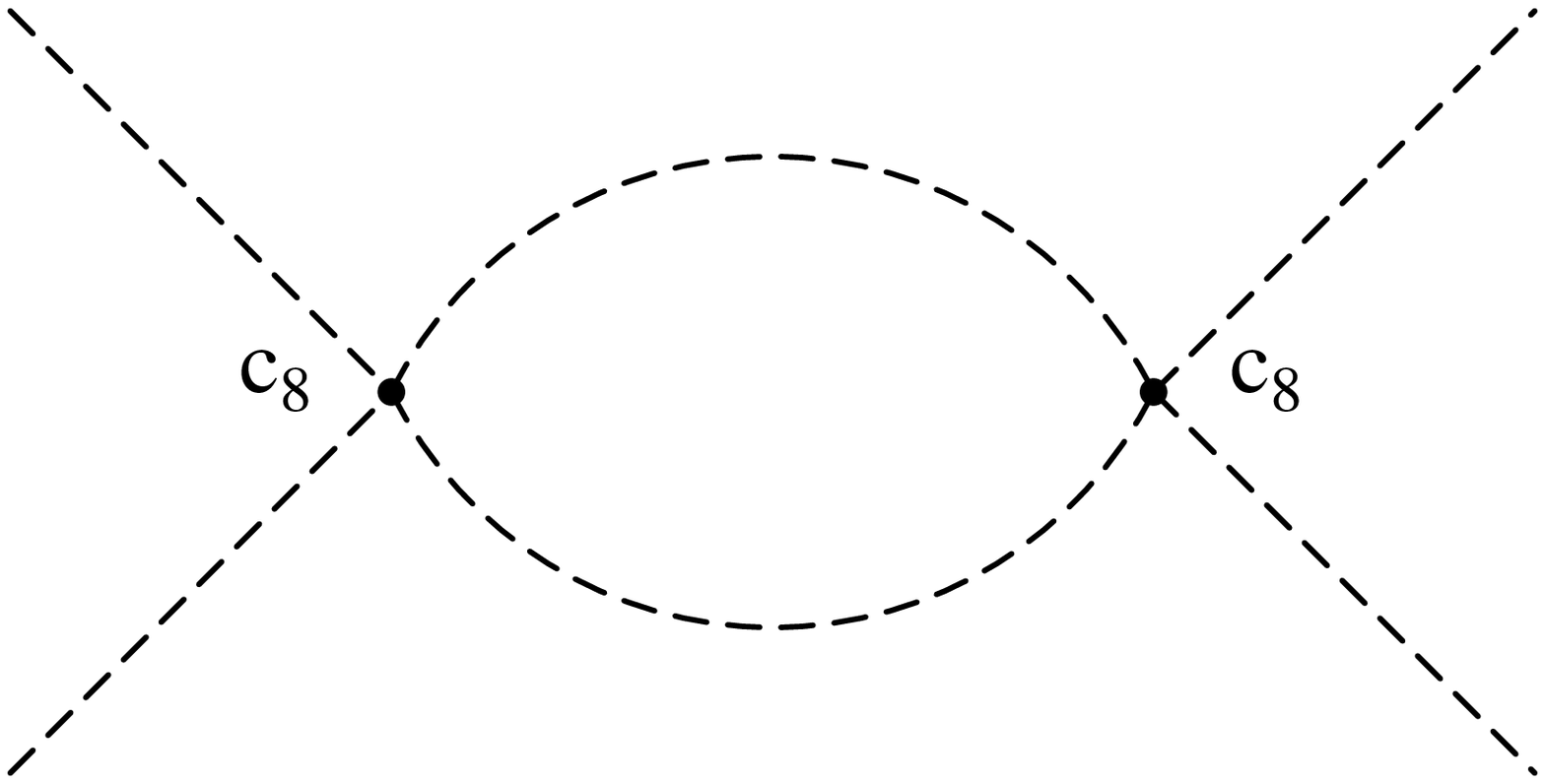} \hfil}\hbox to \size {\hfil(d)\hfil}}}
\caption{Feynman graphs contributing to the $\pi\pi\rightarrow\pi\pi$ in the effective theory.  Graph (a) corresponds to one insertion of the dimension eight operator $(\partial_\mu\pi \partial^\mu\pi)^2$.   Graphs (b), (c) correspond to insertions of a generic dimension ten and twelve operator, respectively.   Note that the $\omega/\Lambda$ power counting scheme implies that the one-loop diagram (d) comes in at the same order as the tree graph (c).  \label{fig:ppEFT}}
\end{figure*}

Given these power counting rules, we now know how to assign powers of $\omega/\Lambda$ to terms in the expansion of any low energy observable.   To see how this works let us consider again $\pi\pi\rightarrow\pi\pi$ scattering.  By the LSZ reduction formula, the $S$-matrix element for this process is of the form
\begin{equation}
S\sim \left[\int d^4 x_i e^{-ik_i \cdot x_i} \partial^2_i \right] \langle 0|T \pi(x_1)\cdots \pi(x_4)|0\rangle.
\end{equation}
Thus to decide the order in which a given operator contributes to the scattering amplitude, we need to know how its contribution to the four-point function $\langle T\pi(x_1)\cdots \pi(x_4)\rangle$ sales in $\omega/\Lambda$.   

For example, the leading non-trivial contribution to scattering, from a single insertion of the dimension eight operator ${\cal O}_8 = (\partial_\mu \pi\partial^\mu\pi)^2$ gives a term whose magnitude is
\begin{eqnarray}
\nonumber
\mbox{Fig.~\ref{fig:ppEFT}(a)} &=& \langle T\pi(x_1)\cdots \pi(x_4)\left[{ic_8\over\Lambda^4}\int d^4 x {{\cal O}_8(x)}\right]\rangle_0\\
 &\sim&  \omega^4 \left[{\omega^{-4}\times \omega^8\over\Lambda^4}\right] = \omega^4 \left({\omega\over\Lambda}\right)^4.
\end{eqnarray}
(The subscript $0$ means we are calculating the correlation functions on the RHS in the free field theory, using Wick's theorem).  At higher orders in the expansion, the four-point function receives corrections from the diagrams in Fig.~\ref{fig:ppEFT}(b), (c), (d).     In particular, a typical contribution from a dimension ten operator ${\cal O}_{10}$ goes like
\begin{eqnarray}
\nonumber
\mbox{Fig.~\ref{fig:ppEFT}(b)}=\langle T\pi(x_1)\cdots \pi(x_4)\left[{ic_{10}\over\Lambda^6}\int d^4 x {\cal O}_{10}(x)\right]\rangle_0 \sim \omega^4 \left({\omega\over\Lambda}\right)^6,\\
\end{eqnarray}
and likewise a single insertion of a dimension twelve operator ${\cal O}_{12}$ (Fig.~\ref{fig:ppEFT}(c)) comes in at relative order $\omega^8/\Lambda^8$.    Note however that at this order there is an additional contribution from one-loop diagrams with two insertions of the dimension eight operator:
\begin{eqnarray}
\nonumber
\mbox{Fig.~\ref{fig:ppEFT}(d)} =  \langle T\pi(x_1)\cdots \pi(x_4)\left[{i c_8\over\Lambda^4}\int d^4 x {\cal O}_8(x)\right]^2\rangle_0\sim \omega^4 \left({\omega\over\Lambda}\right)^8,\\
\end{eqnarray}
thus in practice loop graphs can contribute at the same order as tree level insertions.   

\section{Lecture II}
\label{sec:grEFT}

\subsection{The binary inspiral as an EFT calculation}

\begin{figure}
\includegraphics[width=5cm]{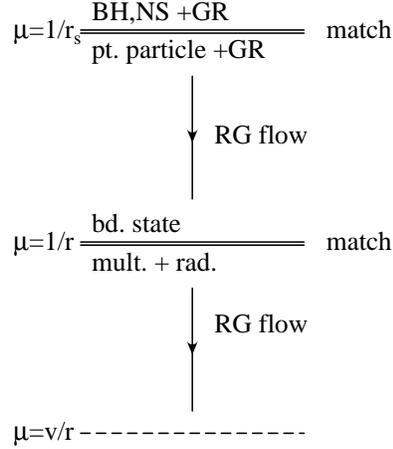}
\caption{Construction of EFT for binary stars in two stages.\label{fig:NRGRschem}}
\end{figure}

As discussed in the previous lecture, there are three scales involved in the binary inspiral problem:   the internal structure scale $r_s$, the orbital distance $r$ and the radiation wavelength $r/v$.     The goal is to calculate gravitational wave observables, e.g. the radiated power, arising from physics at the scale $r/v$.   Following our previous discussion, a convenient way of doing this is to formulate the problem in terms of EFTs.   Since the problem has two intermediate scales, the construction of the EFT that describes the radiation modes must proceed in stages, as shown in Fig.~\ref{fig:NRGRschem}.

The starting point is the theory of an isolated black hole or neutron star coupled to gravity.   Thus in this theory, the relevant degrees of freedom are field perturbations (gravitational, electromagnetic) propagating in the BH/NS background geometry.   Of these degrees of freedom, the field modes of interest in the non-relativistic binary dynamics have wavelengths ($\lambda\sim r/v$) much larger than the scale that characterizes the internal structure of the compact object (e.g., the Schwarzschild radius $r_s$).   Given our previous discussion, it therefore makes sense to integrate out the internal structure of the object by matching onto a new theory that captures the relevant degrees of freedom.   This theory is simply the theory of a point particle coupled to the gravitational  fields (plus whatever other massless fields there are in the problem).   We describe how to construct this EFT (identifying which modes to keep and constructing the Lagrangian) in sec.~\ref{sec:ppEFT}.

This point particle EFT is the correct theory for length scales all the way to the orbital radius $r$.    At the orbital scale, the gravitational field can be viewed as a superposition of ``potential'' modes that act over short distances, mediating the forces that form the bound system, and ``radiation modes'' which correspond to the gravitational waves that propagate out to the detector.   It is convenient once again to integrate out the potential modes by matching onto an EFT of radiation gravitons coupled to a composite object of size $r\ll\lambda$.   This new EFT consists of a point particle, together with a set of multipole mass moments generated by the mechanical plus gravitational energy of the two-particle bound state.   These multipoles are coupled to gravity in a way consistent with gauge invariance under long wavelength coordinate transformations.   We discuss the matching onto, and power counting within this theory in sec.~\ref{sec:NRGR}.   

Note that in Fig.~\ref{fig:NRGRschem}, we have also indicated that the parameters in each of the EFTs relevant to the binary exhibit RG flows as a function of a renormalization scale.   This RG flow, which is purely classical, unfortunately does not arise until order $v^6$ in the binary dynamics, and is beyond the scope of the discussion in these lectures.

\subsection{The EFT for isolated compact objects}
\label{sec:ppEFT}

An isolated compact object probed by long wavelength fields can be approximated as a point particle.   Even if we do not know the internal structure of the compact object, it is possible to write down an EFT that describes its interactions with external fields which accounts for finite size effects in a \emph{systematic} fashion.

Recall from our general discussion that to build an EFT, one needs to do two things:
\begin{itemize}
\item Identify the relevant degrees of freedom at the scale of interest.
\item Construct the most general Lagrangian for these degrees of freedom that is consistent with the symmetries.
\end{itemize} 

A black hole\footnote{Neutron stars may contain additional low frequency modes that must be kept in the point particle EFT.   See~\cite{QNM} for a review of NS/BH spectroscopy.} interacting with long wavelength gravitational fields can be viewed to a first approximation as a point particle probe of the background geometry.  The degrees of freedom necessary to describe such a system are
\begin{enumerate}
\item The gravitational field $g_{\mu\nu}(x)$.
\item The black hole's worldline coordinate $x^\mu(\lambda)$, which is a function of an arbitrary affine parameter $\lambda$.
\item An orthonormal frame $e_a^\mu(\lambda)$ localized on the particle worldline.  This describes the orientation of the object relative to local inertial frames.   It encodes how the particle is spinning relative to the gravitational field.
\end{enumerate}
For the sake of simplicity, I will ignore the effects of spin in the subsequent discussion.  The construction of effective Lagrangians involving spin degrees of freedom can be found in~\cite{spin}.

These degrees of freedom must couple in all ways allowed by the symmetries of the problem.   These are
\begin{enumerate}
\item General coordinate invariance,  $x^\mu\rightarrow x^{\bar\mu}(x)$.
\item Worldline reparametrization invariance (RPI), $\lambda\rightarrow {\bar\lambda}(\lambda)$.
\end{enumerate}
I will make one additional symmetry assumption, again just to keep the discussion as simple as possible:
\begin{enumerate}
\setcounter{enumi}{2}
\item $SO(3)$ invariance.   This guarantees that the compact object is perfectly spherical.  In particular it has no permanent moments relative to its own rest frame.
\end{enumerate}
This last assumption, together with the omission of spin degrees of freedom, means that the EFT that we are constructing is the appropriate one for describing  Schwarzschild black holes interacting with external gravitational fields.   

It is straightforward  to write down effective Lagrangians that are invariant under these symmetries.   To take care of coordinate invariance, we just write down Lagrangians that transform as coordinate scalars constructed from $g_{\mu\nu}(x)$ and $dx^\mu/d\lambda$.   A simple way of ensuring RPI is to use the proper time variable
\begin{equation}
d\tau^2 = g_{\mu\nu}(x(\lambda)) dx^\mu dx^\nu,
\end{equation}
as the worldline parameter.    Since proper time is physical (i.e., measurable) it must be invariant under worldline reparametrizations.

The effective action consistent with these criteria is then
\begin{equation}
\label{eq:PEFT}
S_{eff}[x^\mu, g_{\mu\nu}] = S_{EH}[g]+S_{pp}[x,g],
\end{equation}
where we take the usual action for the gravitational field\footnote{In addition to the Einstein-Hilbert term, the action for gravity may contain additional powers of the curvature suppressed by powers of the scale $m_{Pl}$ where new physics is expected to come in.   These play no role in our discussion.}
\begin{equation}
S_{EH}=- 2 m^2_{Pl}\int d^4 x \sqrt{g} R(x),
\end{equation}
with $m^{-2}_{Pl}=32\pi G_N,$ and $R(x)$ is the Ricci scalar.   $S_{pp}$ is given by
\begin{equation}
S_{pp}=-m\int d\tau + \cdots, 
\end{equation}  
where $m$ is the particle mass, and we have suppressed temporarily any possible curvature dependent terms in the point particle action.    Extremizing $S_{pp}$ gives rise to the usual geodesic motion of a test particle in a gravitational field 
\begin{equation}
\delta \left[-m\int d\tau\right] =0\Longrightarrow {\ddot x}^\mu + {\Gamma^\mu}_{\alpha\beta} {\dot x}^\alpha {\dot x}^\beta \equiv {\dot x}^\alpha D_\alpha  {\dot x}^\mu = 0,
\end{equation}
(${\dot x}^\mu \equiv dx^\mu/d\tau$).  Including only one power of the curvature, $S_{pp}$ has two additional terms
\begin{equation}
S_{pp}=-m\int d\tau + c_R\int d\tau R + c_V\int d\tau R_{\mu\nu} {\dot x}^\mu {\dot x}^\nu \cdots.
\end{equation}
Note however that the leading equations of motion for the gravitational field, which follow varying  $S_{EH}$ and neglecting sources imply that $R_{\mu\nu}(x)=0$.   Thus operators constructed from the Ricci curvature are ``redundant'' operators, and can be omitted from $S_{pp}$ without affecting the physical consequences of the theory.   Technically, this is because it is possible to perform a field redefinition of $g_{\mu\nu}(x)$ which sets the coefficients of these operators to zero.    See the appendix for details.

Since terms with $R_{\mu\nu}$ can be omitted, all that is left are operators constructed from the Riemann tensor (specifically its traceless part, the Weyl tensor).    The simplest such operators involve two powers of the curvature and can be written as
\begin{equation}
\label{eq:finsize}
S_{pp}=-m\int d\tau + c_E\int d\tau E_{\mu\nu} E^{\mu\nu} + c_B\int d\tau B_{\mu\nu} B^{\mu\nu}+\cdots.
\end{equation}
The tensors $E_{\mu\nu}$, $B_{\mu\nu}$ denote the decomposition of the Riemann tensor $R_{\mu\nu\alpha\beta}$ into components of electric and magnetic type parity respectively.   They are the gravitational analog of the decomposition of the electromagnetic field strength $F_{\mu\nu}$ into electric and magnetic fields.   Explicitly, they are given by
\begin{eqnarray}
E_{\mu\nu} &=&  R_{\mu\alpha\nu\beta} {\dot x}^\alpha {\dot x}^\beta,\\
B_{\mu\nu}  &=& \epsilon_{\mu\alpha\beta\rho} {\dot x}^\rho {R^{\alpha\beta}}_{\rho\nu} {\dot x}^\rho.
\end{eqnarray}
These tensors are purely spatial in the particle rest frame and for a background with $R_{\mu\nu}=0$ also satisfy ${E^\mu}_\mu={B^\mu}_\mu=0$.      

The operators $\int d\tau E_{\mu\nu} E^{\mu\nu}$ and $\int d\tau B_{\mu\nu} B^{\mu\nu}$ are the first in an infinite series of terms that systematically encode the internal structure of the black hole.   One way to see that these terms describe finite size effects is to calculate their effect on the motion of a particle moving in a background field $g_{\mu\nu}$.   The variation of $S_{pp}$ including the terms quadratic in the curvature gives
\begin{eqnarray}
\nonumber
\delta\left[-m\int d\tau\right] &=& -\delta\left[ c_E\int d\tau E_{\mu\nu} E^{\mu\nu} + c_B\int d\tau B_{\mu\nu} B^{\mu\nu}+\cdots\right]  \\
& & {} \Longrightarrow  {\dot x}^\alpha D_\alpha {\dot x}^\mu \neq 0  .
\end{eqnarray}
In other words, due to the curvature couplings, the particle no longer moves on a geodesic.    However, geodesic deviation implies stretching by tidal forces, which occurs when one considers the motion of extended objects in a gravitational field.    In fact, an explicit matching calculation, described in Sec.~\ref{sec:finsize}, predicts that the coefficients $c_{E,B}\sim m^2_{Pl} r^5_s$, vanishing rapidly as the size of the black hole goes to zero.

The EFT in Eq.~(\ref{eq:PEFT}) describes the dynamics of one extended object in a gravitational field.    To describe the motion of several objects, we simply include a separate point particle action for each
\begin{equation}
\label{eq:nbodypeft}
S_{eff}[x_a,g]=S_{EH}[g]+ \sum_a S^a_{pp}[x_a,g],
\end{equation}
where the index $a$ runs over all the particles moving in the field $g_{\mu\nu}$.

\subsection{Calculating observables}
\label{sec:obs}

In principle, Eq.~(\ref{eq:nbodypeft}) correctly captures the physics of an arbitrary system of extended objects for scales $\mu< 1/r_s$.   It can therefore be used to calculate all observables measured by gravitational wave detectors.

If we decompose the typical gravitational wave signal as
\begin{equation}
h(t) = A(t) \cos\phi(t),
\end{equation}
then interferometric detectors, such as LIGO/VIRGO and LISA, are particularly sensitive to the phase $\phi(t) \sim 2 \int^t d\tau \omega(\tau)$ of the gravitational wave.   This is usually calculated using the ``adiabatic approximation'':   Consider a non-relativistic binary inspiral, and suppose we have calculated to some order in the $v\ll 1$ expansion the quantities
\begin{eqnarray}
\nonumber
E(v) &=& \mbox{mechanical energy of binary},\\
\nonumber
P(v) &=& \mbox{power emitted in gravitational waves},
\end{eqnarray}
as functions of the orbital parameters, such as the velocity $v(t)$, or equivalently the orbital frequency $\omega(t)$.   Energy conservation
\begin{equation}
{d E\over dt} = - P,
\end{equation}
then gives a differential equation that can be used to solve for $\omega(t)$ which in turn gives the frequency of the GW wave signal $\omega_{GW}(t) \simeq 2\omega(t)$ and consequently the phase $\phi(t)$ (the factor of two arises from the helicity-two nature of the graviton).   We implicitly did this calculation to leading order in $v$ in Sec.~\ref{sec:motivation}.   There we found 
\begin{equation}
\phi(t) = 2\int^t_{t_i} d\tau\omega(\tau) = {1\over 16} \left[{1\over v(t)^5}-{1\over v(t_i)^5}\right],
\end{equation}
for equal mass stars in a circular orbit.

Although the observables we are interested in calculating are purely classical, let us pretend for the moment that we are doing quantum field theory.   Write 
\begin{equation}
\label{eq:h}
g_{\mu\nu} = \eta_{\mu\nu} + {h_{\mu\nu}\over m_{Pl}},
\end{equation}
and calculate the functional $S_{\mathit eff}(x_a)$ defined by the path integral
\begin{equation}
\label{eq:PI}
\exp\left[i S_{\mathit eff}(x_a)\right] = \int D h_{\mu\nu}(x) \exp\left[ i S_{EH}(h) + iS_{pp}(h,x_a)\right],
\end{equation}
with the particle worldlines $x_a^\mu(\tau)$ that source $h_{\mu\nu}$  held fixed.   It turns out that $S_{\mathit eff}(x_a)$ is a generating function for the quantities of interest.  In particular, the classical limit of
the variation
\begin{equation}
\delta \left[\mbox{Re} S_{\mathit eff}(x_a)\right]=0
\end{equation}
gives rise to the equations of motion for the worldlines $x_a^\mu(\tau)$.    From this one can derive an energy function $E(v)$ in the usual way.    Furthermore
\begin{equation}
{1\over T} \mbox{Im} S_{\mathit eff}(x_a) = {1\over 2}\int dE d\Omega {d^2\Gamma\over d\Omega dE},
\end{equation}
measures the total \emph{number} of gravitons emitted as worldlines $x_a^\mu(\tau)$ evolve over a time $T\rightarrow\infty$.    Although graviton number is not a well defined observable classically, the power emitted can be obtained by integrating the differential rate $d\Gamma$ over the energy of the emitted graviton
\begin{equation}
P=\mbox{tot. power} = \int dE d\Omega  \, E {d^2\Gamma\over d\Omega dE}.
\end{equation}

To see how the calculation of observables via Eq.~(\ref{eq:PI}) works, we will consider a toy gravity model consisting of a scalar graviton field $\phi(x)$ interacting with several point particles.   The scalar $\phi$ couples to the point particles with strength proportional to mass
\begin{equation}
S= {1\over 2}\int d^4 x \partial_\mu\phi \partial^\mu \phi -\sum_a m_a\int d\tau_a\left[1 + {\phi\over 2 \sqrt{2} m_{Pl}}\right],
\end{equation}
or equivalently
\begin{equation}
S = - \sum_a m_a\int d\tau_a + \int d^4 x \left[{1\over 2} \partial_\mu\phi \partial^\mu \phi + J(x)\phi(x)\right],
\end{equation}
with 
\begin{equation}
\label{eq:ssource}
J(x) \equiv  -\sum_a {m_a \over 2 \sqrt{2} m_{Pl}}\int d\tau_a \delta^4(x-x_a).
\end{equation}
Then the functional that generates the observables in this model is given by 
\begin{equation}
S_{\mathit eff}(x_a) = -\sum_a m_a \int d\tau_a - i\ln Z[J],
\end{equation}
where
\begin{equation}
\label{eq:Z}
Z[J] = \int D\phi(x) \exp\left[ i\int d^4 x {1\over 2} \partial_\mu\phi \partial^\mu \phi + J(x)\phi(x)\right].
\end{equation}
This is a Gaussian integral so it can be easily calculated explicitly.   Up to an irrelevant constant, the result is 
\begin{equation}
\label{eq:gauss}
\ln Z[J]  = -{1\over 2} \int d^4 x d^4 y J(x) D_F(x-y) J(y).
\end{equation}
In the case of real gravity, with non-linear self interactions, the analog of $Z[J]$ will not have such a simple expression.   Nevertheless, the perturbative expansion of the generating function in real gravity has a simple diagrammatic interpretation.  As a warmup to constructing the diagrammatic rules in real gravity, it is useful to recall how Eq.~(\ref{eq:gauss}) is recovered by summing up Feynman diagrams.  Treat the coupling $\int d^4 x J(x)\phi(x)$ perturbatively, and introduce the Feynman rule
\begin{eqnarray}
\nonumber
\Diagram{fA & fA\\
& gv} &=& i \int d^4 x J(x) e^{-ik\cdot x} = -i\sum_a {m_a\over 2 \sqrt{2} m_{Pl}} \int d\tau_a e^{-i k\cdot x_a}.\\
\end{eqnarray}
Then by expanding the interaction term in Eq.~(\ref{eq:Z}), we see that $Z[J]$ has a diagrammatic expansion in terms of ``ladder diagrams'' 
\begin{eqnarray}
\label{eq:ladder}
\nonumber
Z[J] &=& \Diagram{fA & fA\\
& gv\\
fA & fA} + 
\Diagram{fs & fA & fs\\
 & gv & gv\\
 fs & fA & fs} + 
 \Diagram{fs &fA & fA & fs\\
 & gv & gv & gv\\
 fs &fA & fA & fs} + \cdots  \\
 &=& \exp\left(\Diagram{fA & fA\\
& gv\\
fA & fA}\right).
\end{eqnarray}
Note that the intermediate particle lines have no propagators associated with them (they just depict the time evolution of the particle worldlines), so diagrams with multiple ``rungs'' are simply products of the diagram with a single scalar exchange.    Specifically, the diagram with $n$ intermediate graviton propagators is given by 
\begin{equation}
\Diagram{fs & fA\\
& gv\\
fs & fA}
\cdots
\Diagram{fA & fs\\
& gv\\
fA & fs}
=
{1\over n!} \left[{1\over 2} \int d^4 x d^4 y (i J(x)) D_F(x-y) (i J(y))\right]^n,
\end{equation}
where the factor of $1/ 2^n n!$ appearing here is the symmetry factor associated with the diagram.   Summing up contributions from diagrams with any number $n$ of scalar lines, we see that the series in Eq.~(\ref{eq:ladder}) reproduces Eq.~(\ref{eq:gauss}).

Notice that the exponentiation of the diagrams contributing to $Z[J]$ implies that $S_{\mathit eff}(x_a)$ receives a contribution only from diagrams that remain connected after the particle worldlines are stripped off, which in this theory is just the diagram with a single graviton exchange.   The fact that $S_{\mathit eff}(x_a)$ is given by diagrams that do not break up after removing all worldlines remains true in real gravity, where the graphs can have graviton self-interaction vertices.

The effective action in the toy gravity model is thus given by 
\begin{eqnarray}
\label{eq:tseff}
\nonumber
S_{\mathit eff}(x_a) = -\sum_a m_a\int d\tau_a +{i\over 2}\sum_{a,b} {m_a m_b\over 8 m^2_{Pl}} \int d\tau_a d\tau_b  D_F(x_a-x_b).\\
\end{eqnarray}
Eq.~(\ref{eq:tseff}) contains all the information about the classical particle dynamics.   For illustration, let's  evaluate it for the case where the motion is non-relativistic.  In this case the integral 
\begin{equation}
\int d\tau_a d\tau_b D_F(x_a-x_b) = \int d\tau_a d\tau_b \int{d^4 k\over (2\pi)^4} {i\over k^2+i\epsilon} e^{-ik\cdot (x_a-x_b)}
\end{equation}
can be split up into contributions from two regions of momenta:
\begin{itemize}
\item {\bf Potential}:   This is the region corresponding to gravitons with spacelike momenta of the form
\begin{equation}
k^\mu \rightarrow (k^0\sim {v\over r}, {\bf k}\sim {1\over r}),
\end{equation}
where $r$ is the typical separation between the particles and $v\ll 1$ is the typical three-velocity.   Gravitons with this momentum configuration mediate nearly instantaneous exchanges between the point particles.   Note that potential gravitons can never be on-shell, $k^2\neq 0$.  Therefore they never contribute  to the imaginary part of Eq.~(\ref{eq:tseff}), which happens when the $i\epsilon$ term in the propagator becomes important, i.e., when $k^2=0$.
\item {\bf Radiation}:   Radiation gravitons correspond to on-shell modes with 
\begin{equation}
k^\mu \rightarrow (k^0\sim {v\over r},{\bf k}\sim {v\over r}).
\end{equation}
Because these modes satisfy $k^2\simeq 0$, they are responsible for giving rise to the imaginary part of $S_{\mathit eff}(x_a)$ and therefore the radiation that propagates out to the detector.
\end{itemize}

Since only potential modes contribute to the real part of $S_{\mathit eff}(x_a)$, we may calculate in the NR limit by expanding the propagator as
\begin{equation}
{1\over k^2_0 - {\bf k}^2} =-{1\over {\bf k}^2}\left[1 + {k^2_0\over {\bf k}^2}+\cdots\right] =-{1\over {\bf k}^2}\left[1 + {\cal O}(v^2)\right].
\end{equation}
Using 
\begin{equation}
\int {d^4 k\over (2\pi)^4} e^{-i k\cdot x} {1\over {\bf k}^2} = {1\over 4\pi |{\bf x}|} \delta(x^0),
\end{equation}
and expanding $d\tau_a = dx^0\sqrt{1-{\bf v}^2_a}\simeq dx^0\left[1 -{1\over 2}{\bf v}^2_a\right]$, we find to leading order in $v$
\begin{equation}
\mbox{Re} S_{eff}[x_a] = {1\over 2}\sum_a \int dx^0 m_a {\bf v}^2_a  -{1\over 2}\sum_{a,b}\int dx^0 {G_N m_a m_b\over |{\bf x}_a-{\bf x}_b|} + \cdots,
\end{equation}
which is just the Lagrangian for classical NR particles interacting through a Newtonian potential.  Note that the second term contains divergent self-energy contributions whenever $a=b$ in the sum.    These divergences can be absorbed by renormalization into the particle masses.   Formally, they can just be set to zero, by evaluating the momentum integral in dimensional regularization using the formula
\begin{equation}
\int {d^d {\bf k}\over (2\pi)^d} e^{-i{\bf k}\cdot {\bf x}} {1\over ({\bf k}^2)^\alpha} = {1\over (4\pi)^{d/2}} {\Gamma(d/2-\alpha)\over \Gamma(\alpha)} \left({{\bf x}^2\over 4}\right)^{\alpha-d/2},
\end{equation} 
and taking the limit ${\bf x}\rightarrow 0$ before setting $d=3$.

To calculate the imaginary part of $S_{\mathit eff}(x_a)$ we use
\begin{equation}
\mbox{Im} {1\over k^2 + i\epsilon} = - i\pi\delta(k^2),
\end{equation}
which physically has the effect of ensuring that only on-shell particles contribute to the radiated power.  Then 
\begin{equation}
\mbox{Im} S_{eff}(x_a) 
= {1\over 16 m^2_{Pl}}\int {d^3 {\bf k}\over (2\pi)^3} {1\over 2|{\bf k}|}  \left|\sum_a m_a \int d\tau_a e^{-i k\cdot x_a}\right|_{k^0=|{\bf k}|}^2,
\end{equation} 
and the differential power is
\begin{equation}
\label{eq:dpow}
{d P\over d\Omega d{|\bf k|}} = {1\over T} {G_N\over 4\pi^2}  {\bf k}^2\left|\sum_a m_a \int d\tau_a e^{-i k\cdot x_a}\right|_{k^0=|{\bf k}|}^2.
\end{equation}
This is exactly what one would find classically, by solving the wave equation for $\phi$ with source term given by Eq.~(\ref{eq:ssource}) and then calculating its energy-momentum tensor at an asymptotically large distance from the source.   It can also be calculated from the tree-level amplitude for single graviton emission from the source particles
\begin{equation}
i{\cal A}(\mbox{vac.}\rightarrow \phi) = - i \sum {m_a\over 2 m_{Pl} \sqrt{2}}\int d\tau_a e^{i k\cdot x_a}.
\end{equation}
The differential \emph{probability} for the emission of one graviton is then
\begin{equation}
d\mbox{Prob}(\mbox{vac.}\rightarrow\phi) = |{\cal A}(\mbox{vac.}\rightarrow \phi)|^2 {d^3 {\bf k}\over (2\pi)^3} {1\over 2 |{\bf k}|}. 
\end{equation}
Dividing this result by the observation time $T\rightarrow\infty$ to get a rate, and multiplying by the graviton energy $|{\bf k}|$ to convert to power reproduces Eq.~(\ref{eq:dpow}).  So indeed the functional integral formula for $S_{eff}(x_a)$ knows about the classical observables for the particle ensemble.

{\bf Exercise}:   In the case of electrodynamics coupled to point particles, 
\begin{equation}
S= -{1\over 4}\int d^4 x F_{\mu\nu} F^{\mu\nu} -\sum_a m_a\int d\tau_a +\sum_a e Q_a\int dx_a^\mu A_\mu,
\end{equation}
show by calculating $\mbox{Im} S_{eff}(x_a)$ that the radiated power in photons is given by 
\begin{equation}
{dP\over d\Omega d|{\bf k}|}= {1\over T} {\alpha\over 4\pi^2} {\bf k}^2\left|\sum_a Q_a \int dx_a^\mu e^{-ik\cdot x_a}\right|^2_{k^0=|{\bf k}|},
\end{equation}
where $\alpha=e^2/(4\pi)$.   Verify that, upon time averaging, this reproduces the usual electric dipole radiation formula in the non-relativistic limit.

Gravitational radiation in general relativity can be calculated by a straightforward generalization of the methods used for the scalar gravity model.   Real gravity is of course more complex, but the complications are mainly computational.    If we expand $g_{\mu\nu}$ as in Eq.~(\ref{eq:h}) and plug into the gravitational action we find an infinite series of terms that are schematically of the form
\begin{eqnarray}
\nonumber
-2 m^2_{Pl}\int d^4 x \sqrt{g} R(x) &\rightarrow& \int d^4 x \left[(\partial h)^2 + {h (\partial h)^2\over m_{Pl}} + {h^2 (\partial h)^2\over m^2_{Pl}}+\cdots\right],\\
\nonumber
&=&
(\Diagram{g})^{-1}\,\,\,+ 
\Diagram{gd\\
& g\\
gu}\,\,\, +
\Diagram{gd & gu\\
gu & gd}\,\,\, +
\cdots,\\
\end{eqnarray}
leading to graviton self-interactions with Feynman vertices containing any number of graviton lines.  We will not need the detailed form of these terms in what follows.    A derivation of the Feynman rules for gravity can be found in the lectures by Veltman~\cite{veltman}.   See also~\cite{donoghue}.   In addition, the gravitational field has non-linear interactions with the point particle, e.g.,
\begin{eqnarray}
\nonumber
-m\int d{\bar\tau}\sqrt{1+{h_{\mu\nu} {\dot x}^\mu {\dot x}^\nu\over m_{Pl}}} &=&-m\int d{\bar\tau} -{m\over 2 m_{Pl}}\int d{\bar\tau} h_{\mu\nu} {\dot x}^\mu {\dot x}^\nu \\
& & {} - {m\over 8 m^2_{Pl}}\int d{\bar\tau}(h_{\mu\nu} {\dot x}^\mu {\dot x}^\nu)^2 +\cdots,
\end{eqnarray}
where $d{\bar\tau}^2=\eta_{\mu\nu} dx^\mu dx^\nu$.   This gives rise to vertices
\begin{equation}
\Diagram{fvA \\
g\\
fvA} \,\,\, +  \,\,\,\,
\Diagram {fvA\\
fvA}
\Diagram{gu\\
gd} \,\,\, +  \,\,\,\,
\Diagram {fvA\\
\\fvA}
\Diagram{gu \\
g\\gd  }\,\,\,+\cdots. 
\end{equation}

\begin{figure}
\def\size{4cm}
\hbox{\vbox{\hbox to \size {\hfil \includegraphics[width=2.5cm]{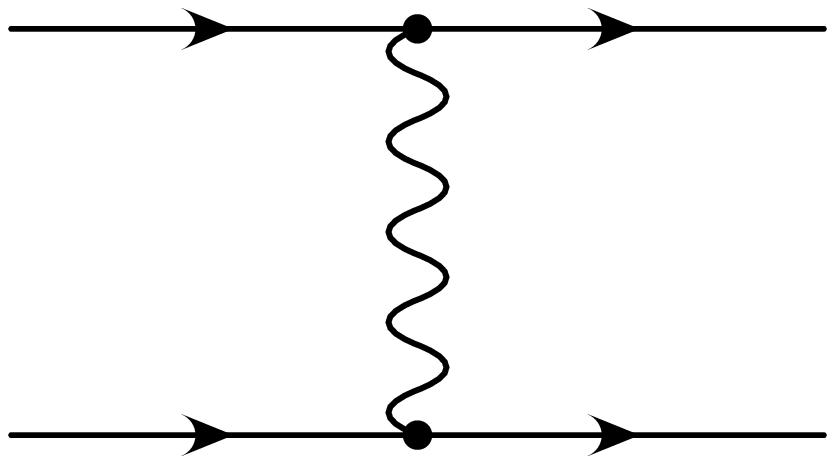} \hfil }\hbox to \size {\hfil(a)\hfil}}
\vbox{\hbox to \size {\hfil \includegraphics[width=2.5cm]{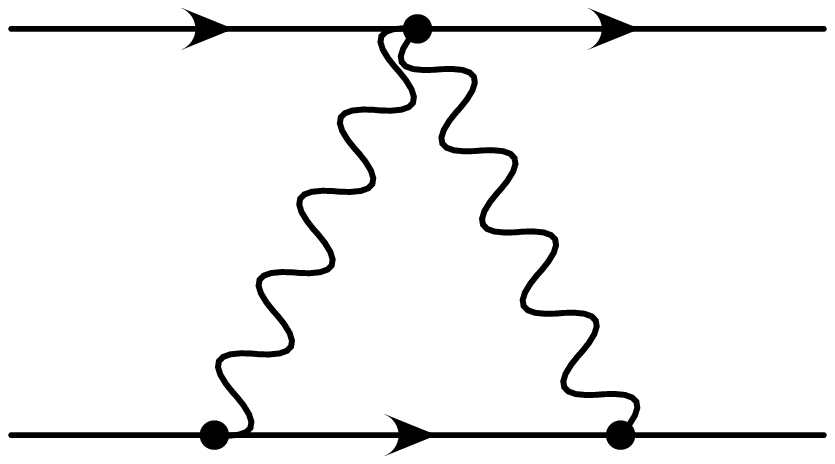} \hfil}\hbox to \size {\hfil(b)\hfil}}
\vbox{\hbox to \size {\hfil \includegraphics[width=2.5cm]{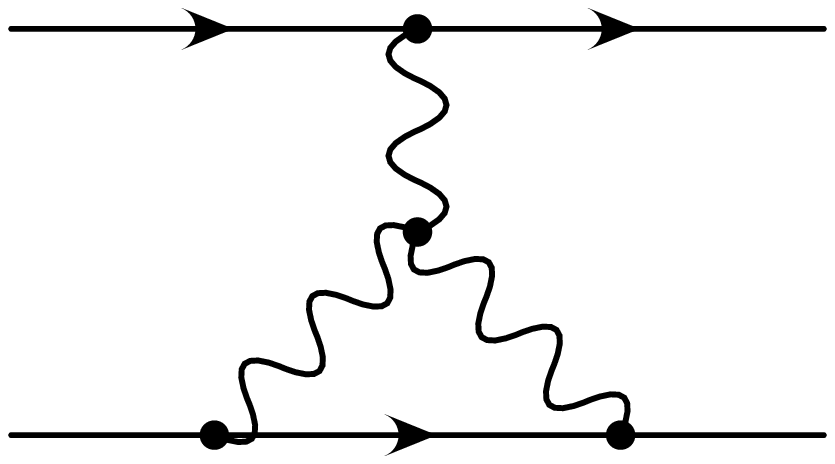} \hfil}\hbox to \size {\hfil(c)\hfil}}}
\caption{The first few diagrams contributing to $S_{\mathit eff}(x_a)$ in Lorentz covariant perturbation theory.\label{fig:relexp}}
\end{figure}

As a result of this, the diagrammatic expansion of $S_{eff}(x_a)$ in the gravitational case is much richer than in the simple model considered above.    The first few terms in the expansion are shown in Fig.~\ref{fig:relexp}.   The problem with this (covariant) form of the perturbative series is that it is not optimal for the calculation of observables in the $v\ll 1$ limit.   For example, consider the diagram in Fig.~\ref{fig:relexp}(c).   At what order in $v$ does  it contribute?     In the language of the previous lecture, we do not have a velocity \emph{power counting} scheme for the diagrams appearing in Fig.~\ref{fig:relexp}.   Since in the end we are interested in computing gravitational wave signals to a fixed order in $v$, it is important to develop a set of rules that assigns a unique power of $v$ to each diagram in the theory.

\subsection{Integrating out the orbital scale}
\label{sec:NRGR}

The reason why the diagrams do not scale as definite powers of $v$ is that there are multiple scales in the theory.   Although we have successfully integrated out the internal structure scale $r_s$, the momentum integrals in the diagrams of Fig.~\ref{fig:relexp} have contributions both from potential $(v/r,1/r)$ and radiation $(v/r,v/r)$ gravitons.   But potential modes are never on-shell (they have $k^2\sim 1/r$) and do not belong in the effective theory at long distance scales.   

To remedy this problem, we will explicitly treat the potential and radiation modes separately, by splitting up the gravitational field as
\begin{equation}
\label{eq:ppr}
h_{\mu\nu}(x)= {\bar h}_{\mu\nu}(x) + H_{\mu\nu}(x).
\end{equation}
Here, the field ${\bar h}_{\mu\nu}$ represents the long-wavelength radiation modes.   Schematically, it satisfies
\begin{equation}
\partial_\alpha {\bar h_{\mu\nu}}\sim {v\over r} {\bar h}_{\mu\nu},
\end{equation}
meaning that the field ${\bar h}_{\mu\nu}$ varies slowly over spacetime, with a typical length scale $r/v$.   It can therefore be regarded as a slowly varying background field in which the potential modes $H_{\mu\nu}$, with 
\begin{eqnarray}
\partial_0 H_{\mu\nu} \sim {v\over r} H_{\mu\nu},  & \partial_i H_{\mu\nu}\sim {1\over r} H_{\mu\nu},
\end{eqnarray}
propagate.   Actually it is useful to re-write $H_{\mu\nu}$ in terms of its Fourier transform
\begin{equation}
\label{eq:FT}
H_{\mu\nu}(x)=\int {d^3{\bf k}\over(2\pi)^3} e^{i{\bf k}\cdot {\bf x}} H_{{\bf k}\mu\nu}(x_0).
\end{equation}
This allows us to explicitly disentangle hard momenta 
\begin{equation}
{\bf k}\sim {1/r},
\end{equation}
from the long wavelength scale 
\begin{equation}
x^\mu\sim {r\over v},
\end{equation}
and to treat all derivatives acting on fields on the same footing, $\partial_\mu\sim v/r$.

To derive an EFT that has manifest velocity power counting rules, we need to integrate out the potential modes.    The basic idea is the following:  we calculate $S_{\mathit eff}(x_a)$ (and hence all observables), in two stages.   In the first stage, we perform the integral over $H_{{\bf k}\mu\nu}$, 
\begin{eqnarray}
\label{eq:NRGRint}
e^{i S_{NR}({\bar h},x_a)} =\int D H_{{\bf k},\mu\nu}(x^0) e^{i S_{EH}({\bar h}+H) + S_{pp}({\bar h}+H,x_a)}
\end{eqnarray}
where the field ${\bar h}_{\mu\nu}$ is treated as a background field.    This gives the quantity $S_{NR}({\bar h}, x_a),$ which formally contains the two-body forces between the point particles, written as an explicit expansion in powers of $v$, and the couplings of the particles to radiation.    In this theory, the short distance scale $r$ appears explicitly, in the coefficients of operators.   In other words, Eq.~(\ref{eq:NRGRint}) is simply a formal way of doing the \emph{matching} to the long-distance EFT containing ${\bar h}_{\mu\nu}$ and the particle worldlines.      Given $S_{NR}({\bar h},x)$, the second stage in the calculation is to perform the functional integral over ${\bar h}_{\mu\nu}$.   Feynman diagrams are easier to calculate in this theory than in the full theory, as momentum integrals only contain the scale $r/v$.

Note that in Eq.~(\ref{eq:NRGRint}) we have  dropped gauge fixing terms necessary to make sense of the path integral.    We will not need the explicit form of such terms in the subsequent discussion, but it is perhaps worth noting that it is very convenient to do the gauge fixing in a way that preserves the invariance under diffeomorphisms that transform the background metric ${\bar h}_{\mu\nu}$ (this is called the background field method~\cite{background}).   If such a gauge fixing scheme is chosen, the action for ${\bar h}_{\mu\nu}$ is guaranteed to be gauge invariant.  This places strong restrictions on the form of the EFT that describes radiation.

Diagrammatically, $S_{NR}({\bar h},x_a)$ is given by a sum over diagrams that have the following topological properties:
\begin{enumerate}
\item Diagrams must remain connected if the particle worldlines are stripped off.
\item Diagrams may \emph{only} contain internal lines corresponding to propagators for the potential modes $H_{{\bf k}\mu\nu}(x_0)$.   Diagrams \emph{cannot} contain external potential graviton lines.
\item Diagrams may \emph{only} contain external ${\bar h}_{\mu\nu}$ lines.    Since the functional integral in Eq.~(\ref{eq:NRGRint}) views the field ${\bar h}_{\mu\nu}$ as a background, diagrams \emph{cannot} contain propagators corresponding to internal radiation graviton lines.
\end{enumerate}

The point of splitting the original graviton $h_{\mu\nu}$ into the new modes ${\bar h}_{\mu\nu}$, $H_{{\bf k}\mu\nu}$ is that the diagrams in terms of these new variables can be assigned definite powers of the expansion parameter $v$.   The power counting rules for determining how many powers of $v$ to assign to a given diagram follow simply from the fact that the three momentum of a potential graviton scales as ${\bf k}\sim 1/r$, since this is the range of the force it mediates, and that the spacetime variation of a radiation graviton is $x^\mu\sim r/v$.   These two observations are sufficient to consistently assign powers of $v$ to any term in the action, and by extension to the Feynman rules.

First, let's determine the power counting for diagrams that only contain potential modes.    We can fix the scaling of $H_{\bf k\mu\nu}$ in terms of kinematic variables by looking at its propagator.  To obtain the propagator, we need the quadratic terms in $H_{\bf k\mu\nu}$ from the gravitational Lagrangian.   Including an appropriate gauge fixing term, whose form is not necessary for our purposes here (see~\cite{GnR1} for details) the relevant terms are
\begin{equation}
S_{H^2} = -{1\over 2}\int dx^0 {d^3 {\bf k}\over (2\pi)^3} \left[{\bf k}^2 H_{{\bf k}\mu\nu} H^{\mu\nu}_{\bf -k} -{{\bf k}^2\over 2} H_{\bf k} H_{\bf -k}\right],
\end{equation}
where $H_{\bf k} = {H^\alpha}_{\alpha{\bf k}}$.  Terms with time derivatives have been dropped, as $\partial_0$ is down relative to ${\bf k}$ by a power of $v$.    Because of this, the potential propagator is instantaneous,
\begin{equation}
\label{eq:pprop}
\langle T H_{\bf k\mu\nu}(x^0) H_{\bf q\alpha\beta}(0)\rangle = -{i\over{\bf k}^2} (2\pi)^3 \delta^3({\bf k}+{\bf q})\delta(x^0) P_{\mu\nu;\alpha\beta},
\end{equation}
where the tensor structure is given by 
\begin{equation}
P_{\mu\nu;\alpha\beta}={1\over 2}\left[\eta_{\alpha\mu}\eta_{\beta\nu}+\eta_{\alpha\nu}\eta_{\beta\mu}-{\eta}_{\mu\nu}\eta_{\alpha\beta}\right].
\end{equation}
Eq.~(\ref{eq:pprop}) is essentially the Fourier transform of the Newton potential, as one would expect.  Since  $P_{\mu\nu;\alpha\beta}$ is ${\cal O}(1)$, the scaling of $H_{\bf k\mu\nu}$ is simply
\begin{equation}
H^2_{\bf k\mu\nu}\sim \left({1\over r}\right)^{-2} \times \left({1\over r}\right)^{-3}\times \left({r\over v}\right)^{-1} = r^4 v,
\end{equation}
so that 
\begin{equation}
H_{\bf k\mu\nu}\sim r^2 \sqrt{v}.  
\end{equation}
Given the scaling rule for $H_{\bf k\mu\nu}$, it is now possible to assign powers of velocity to diagrams with no external radiation gravitons.   For example, the coupling of $H_{{\bf k} 00}$ to a NR particle is 
\begin{eqnarray}
\label{eq:nvertex}
\nonumber
-{m\over 2 m_{Pl}}\int dx^0 {d^3 {\bf k}\over (2\pi)^3} e^{i{\bf k}\cdot {\bf x}(x^0)} H_{{\bf k}00}  &\sim& {m\over m_{Pl}} \times \left({r\over v}\right) \times \left({1\over r}\right)^3 \times \left (r^2 v^{1/2}\right) \\
\nonumber
&=& v^{-1/2} {m\over m_{Pl}}.\\
\end{eqnarray}
The scaling of $m/m_{Pl}$ is fixed by the virial theorem,
\begin{equation}
v^2\sim {G_N m\over r}\Rightarrow {m^2\over m^2_{Pl}} \sim m v^2 r  = L v,
\end{equation}
where we have used $G_N\sim 1/m^2_{Pl},$ and have introduced the orbital angular momentum $L=m vr$.   The interaction in Eq.~(\ref{eq:nvertex}) then scales as  
\begin{equation}
\Diagram{fvA\\
h\\
fvA} =  
-{m\over 2 m_{Pl}}\int dx^0 {d^3 {\bf k}\over (2\pi)^3} e^{i{\bf k}\cdot {\bf x}(x^0)} H_{{\bf k}00}\sim L^{1/2},
\end{equation}
so the exchange diagram with two insertions of this term, which leads to the Newton potential between the point particles, scales as a power of the angular momentum $L$,
\begin{eqnarray}
\Diagram{fvA & fvA\\
h\\
fvA & fvA} &\sim& L.
\end{eqnarray}
It can be shown that any diagram without external radiation lines scales like $L^n v^m$, with $n\leq 1$ and $m\geq 0$.   The bound on $n$ is saturated by diagrams without graviton loops.   Thus $L$ is the loop counting parameter of this theory, and since we are interested in the classical limit $\hbar\rightarrow 0$, it is a good approximation to drop diagrams with graviton loops (formally, such diagrams are down by powers of $\hbar/L\ll 1$).

\begin{figure*}
\def\size{5cm}
\hbox{\vbox{\hbox to \size {\hfil \includegraphics[width=2.5cm]{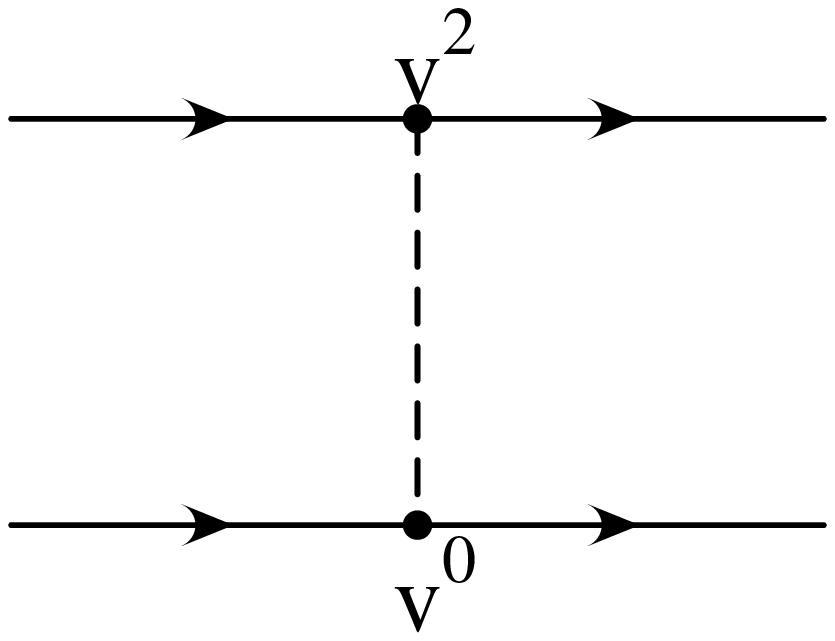} \hfil }\hbox to \size {\hfil(a)\hfil}}
\vbox{\hbox to \size {\hfil \includegraphics[width=2.5cm]{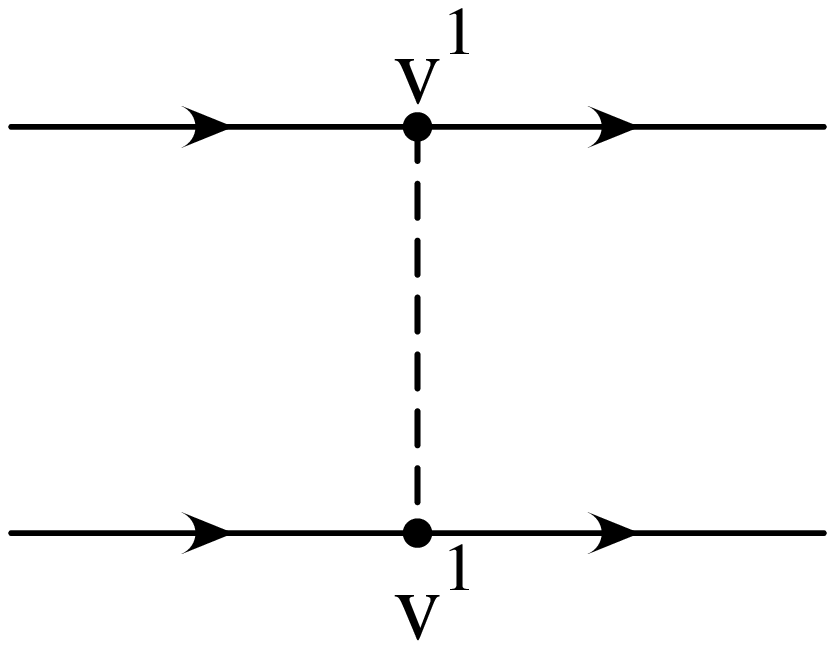} \hfil}\hbox to \size {\hfil(b)\hfil}}}
\vspace{0.25cm}
\hbox{\vbox{\hbox to \size {\hfil \includegraphics[width=2.5cm]{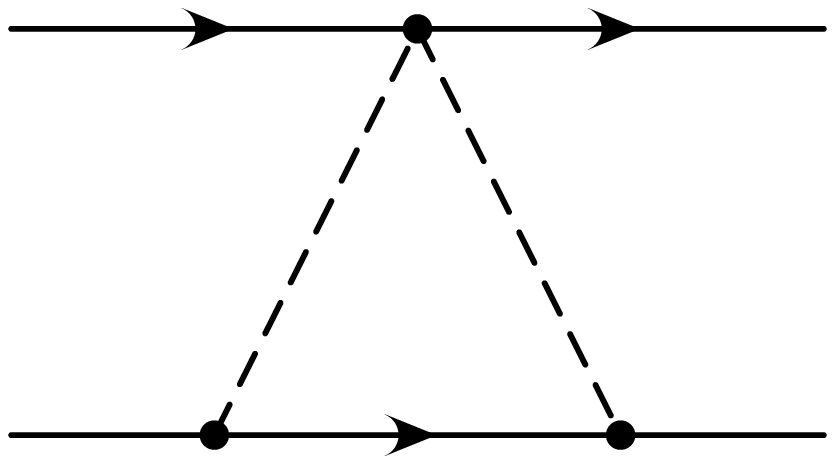} \hfil }\hbox to \size {\hfil(c)\hfil}}
\vbox{\hbox to \size {\hfil \includegraphics[width=2.5cm]{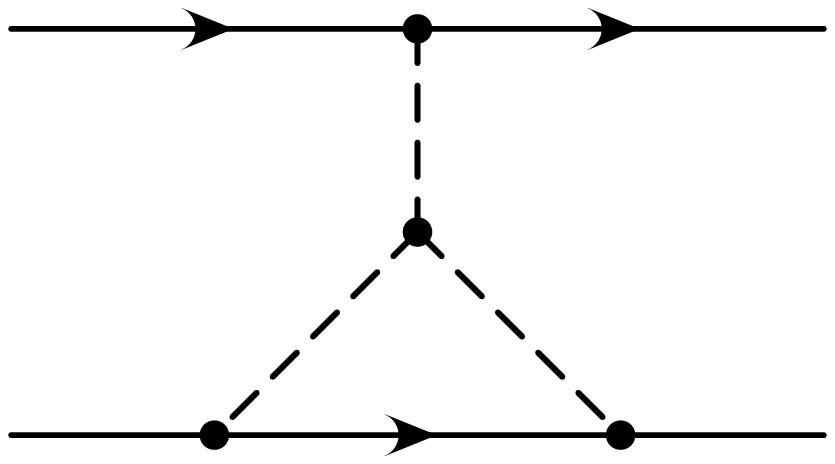} \hfil}\hbox to \size {\hfil(d)\hfil}}}
\caption{Diagrams contributing to the order $v^2$ corrections to the two-body potential.  \label{fig:1PN}}
\end{figure*}

Power counting rules for potential graviton self-interactions can be derived along similar lines.   Consider the potential three-graviton vertex.   Expanding the gravitational Lagrangian to cubic order in $H_{{\bf k}\mu\nu}$, one finds terms with the structure
\begin{equation}
S_{H^3} \sim {1\over m_{Pl}} \int dx^0 (2\pi)^3\delta^3(\sum_i {\bf k}_i) {\bf k}^2 \prod_{i=1}^3 {d^3 {\bf k}_i\over (2\pi)^3}  H_{{\bf k}_i},
\end{equation}
where we have inserted the expansion Eq.~(\ref{eq:FT}) into the Einstein-Hilbert action and performed the $d^3 {\bf x}$ integral .   The factor of ${\bf k}^2$ comes from the fact that every term in the action has two derivatives on the metric.   From this we find
\begin{eqnarray}
\nonumber
\Diagram{hd\\ & h \\ hu}
&\sim& {1\over m_{Pl}}\times\left({r\over v}\right)\times \left({1\over r}\right)^{-3}\times\left({1\over r}\right)^2\times\left[\left({1\over r^3}\right)\times \left(r^2\sqrt{v}\right)\right]^3\\
&=&  {v^2\over\sqrt{L}},
\end{eqnarray}
and therefore the diagram 
\begin{equation}
\Diagram{fvA \\fvA}\Diagram{hd\\ & h \\ hu}\Diagram{fvA\\ fvA}\sim \left(\sqrt{L}\right)^2\times {v^2\over \sqrt{L}}\times \sqrt{L} = L v^2,
\end{equation}
gives rise to a term in the two-body potential that is suppressed by $v^2$ relative to the leading order Newtonian exchange diagram.   The full set of potentials at order $v^2$ arise from the diagrams in Fig.~\ref{fig:1PN}.  In addition to the three graviton term just discussed, there are additional velocity suppressed vertices from the expansion of the particle proper time Lagrangian in powers of $v$,
\begin{equation}
S_{pp}\Rightarrow {m\over m_{Pl}}\int dx^0\left[-{1\over 2} h_{00} -h_{0i} v_i -{1\over 4} h_{00} {\bf v}^2 -{1\over 2} h_{ij} v^i v^j\right].
\end{equation}
The sum of these diagrams gives the corrections to the two-body NR Lagrangian at order $v^2$
\begin{eqnarray}
\nonumber
L_{v^2} &=& {1\over 8}\sum_a m_a {\bf v}^4_a + {G_N m_1 m_2\over |{\bf x}_1-{\bf x}_2|}\left[3({\bf v}^2_1 + {\bf v}^2_2) - 7 {\bf v}_1\cdot {\bf v}_2 \right.\\
& & {}\left. -{({\bf v}_1\cdot {\bf x}_{12})({\bf v}_2\cdot {\bf x}_{12})\over |{\bf x}_1-{\bf x}_2|}\right] - {G^2_N m_1 m_2 (m_1+m_2)\over 2 |{\bf x}_1-{\bf x}_2|^2},
\end{eqnarray}
where ${\bf x}_{12}={\bf x}_1-{\bf x}_2$.   This Lagrangian was first obtained by Einstein, Infeld and Hoffman~\cite{EIH} in 1938.    In this equation, the first term is just the first relativistic correction to the particle kinetic energies, the second term arises from diagrams with a single graviton exchanged by velocity-dependent vertices, and the last term is from the diagrams of Fig.~\ref{fig:1PN}(c), (d).   Diagrams at higher orders in $v$ can be calculated and power counted by the same methods outlined here.

\subsection{Radiation}

Integrating out the non-dynamical potential modes generates the gravitational forces between the non-relativistic particles.   These arise from diagrams with no external factors of the radiation field ${\bar h}_{\mu\nu}$.    The functional integral in Eq.~(\ref{eq:NRGRint}) also generates, from the diagrams with one or more external radiation graviton, the couplings of matter to radiation.    

Incorporating radiation is fairly straightforward.   First note that the propagator for the radiation field, in a suitable gauge, is given by 
\begin{equation}
\label{eq:radprop}
\langle T {\bar h}_{\mu\nu}(x) {\bar h}_{\alpha\beta}(0)\rangle = \int {d^4 k\over (2\pi)^4} {i\over k^2+i\epsilon} e^{-ik\cdot x} P_{\mu\nu;\alpha\beta}.
\end{equation}
Since $k^\mu\sim v/r$, Eq.~({\ref{eq:radprop}) implies that radiation modes should scale as
\begin{equation}
{\bar h}_{\mu\nu}\sim {v\over r}.
\end{equation}
This rule allows one to power count terms in the action containing the radiation field.   However, in order to obtain an EFT for radiation that has manifest velocity power counting, the decomposition of the graviton into potential and radiation modes is not sufficient.    It is necessary also to multipole expand the   couplings of the radiation field to either potentials or to the point particles,
\begin{eqnarray}
\label{eq:mult}
\nonumber
{\bar h}_{\mu\nu}({\bf x},x^0) &=&  {\bar h}_{\mu\nu}({\bf X},x^0) + \delta {\bf x}^i \partial_i {\bar h}_{\mu\nu}({\bf X},x^0) + {1\over 2} \delta {\bf x}^i  \delta {\bf x}^j {\bar h}_{\mu\nu}({\bf X},x^0)\\
& & {}\, +\cdots,
\end{eqnarray} 
where ${\bf X}$ is an arbitrary point, for example the center of mass of the multi-particle system, ${\bf X}_{cm}=\sum_a m_a{\bf x}_a/\sum_a m_a$.   The reason for this is familiar in the case of couplings to matter sources:   consider the amplitude for graviton emission by an ensemble of point particles ($v_a^\mu$ is the four-velocity, $\epsilon_{\mu\nu}(k)$ is the graviton polarization)
\begin{equation}
i{\cal A} = \sum_a\maxis{\Diagram{& gu\\ fA & fA}}= -i\sum_a {m_a\over 2m_{Pl}}\int d\tau_a e^{ik\cdot x_a}v_a^\mu v_a^\nu  \epsilon^*_{\mu\nu}(k).
\end{equation}
The final state graviton is on-shell, so $k^0=|{\bf k}|$.  On the other hand, if the particles are non-relativistic, $x^0\gg |{\bf x}_a|$ (measuring positions relative to the center of mass).  Then
\begin{eqnarray}
\nonumber
e^{-ik\cdot x_a} = e^{-ik_0 x^0} e^{i{\bf k}\cdot {\bf x}_a} =e^{-ik_0 x^0}\left[1+ i{\bf k}\cdot {\bf x}_a -{1\over 2} ({\bf k}\cdot {\bf x}_a)^2 +\cdots\right].\\
\end{eqnarray}
Note that if $|{\bf k}|\sim v/r$ and $|{\bf x}_a|\sim r$, ${\bf k}\cdot {\bf x}_a\sim v$.   So the exponential contains an infinite number of powers of $v$, and the amplitude does not scale homogeneously in velocity.    This is avoided by multipole expanding ${\bar h}_{\mu\nu}$ at the level of the Lagrangian.
Likewise, if we consider a graph containing potential as well as radiation modes
\begin{eqnarray}
\mbox{Fig.~\ref{fig:mult}}\sim{1\over ({\bf p}+{\bf k})^2} = {1\over {\bf p}^2}\left[1 -2 {\bf p}\cdot {\bf k}+\cdots \right],
\end{eqnarray}
we find that because $|{\bf p}|\sim 1/r$ and $|{\bf k}|\sim v/r,$ the propagator for the potential mode with three-momentum ${\bf p}+{\bf k}$ contains an infinite number of powers of $v$.   To ensure that this does not happen, it is necessary to arrange that radiation gravitons do not impart momentum to the potential modes.   This is exactly what occurs if we plug in the expansion Eq.~(\ref{eq:mult}) into the couplings generated by the gravitational Lagrangian.   The necessity to multipole expand potential-radiation couplings was first pointed out in the context of non-relativistic gauge theories in ref.~\cite{QCDmult}.
\begin{figure}
\includegraphics[width=5cm]{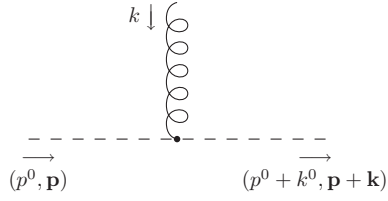} 
\caption{The interaction of a potential and a radiation mode.\label{fig:mult}}
\end{figure}

Given these ingredients, we are finally in a position to calculate the matching coefficients for the radiation graviton EFT at the scale $\mu\simeq 1/r$.   Matching boils down to calculating Eq.~(\ref{eq:NRGRint}), or equivalently comparing diagrams with any number of external radiation modes in the full theory and the EFT.    For example, consider the graphs with one external ${\bar h}_{\mu\nu}$.   In the theory with both potentials and radiation, the leading one is
\begin{eqnarray}
\label{eq:lv0}
\sum_a
\Diagram{gd & ![lrt]{fvA}{v^0} \\
                        & fvA              }
\,\,\,\,\,\Rightarrow                        
\Diagram{gd & {fvA}\, ![lrt]{fv}{v^0}\\
                        & {fvA}\, fv}
&=&  -{1\over 2 m_{pl}}\int dx^0 \sum_a m_a {\bar h}_{00}(0,x^0).
\end{eqnarray}
In this equation, the double lines on the graph on the right indicates that this is a vertex in the EFT below $\mu\simeq 1/r$, where the separation $r$ between the point particles (a short distance scale) cannot be resolved.    It is the vertex obtained by matching at the potential scale, and it corresponds to the term in the EFT shown on the right hand side of the equation.   At this order, the matching is just the zeroth order multipole expansion (performed about the center of mass), i.e. which generates the coupling of ${\bar h}_{00}$ to the mass monopole $m=\sum_a m_a$.   At the next order, the effective theory vertex contains terms both from the multipole expansion and from the explicit velocity dependence of the point particle gravitational couplings.  The result is
\begin{eqnarray}
\sum_a
\Diagram{gd & ![lrt]{fvA}{v^1} \\
                        & fvA              }
\,\,\,\,\,\Rightarrow                        
\Diagram{gd & {fvA}\, ![lrt]{fv}{v^1}\\
                        & {fvA}\, fv}
&=& -{1\over 2 m_{pl}}\int dx^0\left[ {\bf X}^i_{cm}\partial_i {\bar h}_{00} + 2 {\bf P}^i {\bar h}_{0i}\right],
\end{eqnarray}
where ${\bf P}=\sum_a m_a {\dot{\bf x}}_a$ is the total linear momentum of the system.  In the CM frame ${\bf X}_{cm}={\bf P}_{cm}=0$, so this coupling vanishes, i.e., there is no dipole radiation in general relativity.

Things are more interesting in the one-graviton sector at order $v^2$.   The graphs at this order are
\begin{equation}
\label{eq:N2LOrad}
\sum_a
\Diagram{gd & ![lrt]{fvA}{v^2} \\
                        & fvA              }
\,\,\,\,\,\,\,+
\Diagram{gd & fvA & & fvA\\
                        & hs & hs &  \\
                        &fvA & & fvA}   \,\,\,+\,\,\,
\Diagram{fvA & hs & gv & hs & fvA\\
fvA& & & & fvA}
\Rightarrow \Diagram{gd & {fvA}\, ![lrt]{fv}{v^2}\\
                        & {fvA}\, fv}
\end{equation}
where the first graph depicts the terms from the multipole expansion and from the point-particle couplings containing two powers of the velocity, the second graph (an almost identical mirror image graph not shown) comes from the two-graviton couplings of the point particle, and the vertex in the third term is from the ${\bar h} H^2$ terms in the Einstein-Hilbert Lagrangian.  The sum of these graphs gives 
the second order terms in the EFT, $S_{v^2}[{\bar h}]=\int dx^0 L_{v^2}[{\bar h}]$, where 
\begin{eqnarray}
\label{eq:LV2}
\nonumber
L_{v^2}[{\bar h}] &=& -{1\over 2 m_{Pl}} {\bar h}_{00}\left[{1\over 2}\sum_a m_a {\bf v}^2_a - {G_N m_1 m_2\over |{\bf x}_1-{\bf x}_2|}\right]  -{1\over 2 m_{Pl}} \epsilon_{ijk} {\bf L}_k \partial_j {\bar h}_{0i}\\
 & & {} +{1\over 2 m_{Pl}}\sum_a m_a {\bf x}^i_a {\bf x}^j_a R_{0i0j}.
\end{eqnarray}
Here,  the first term is just the coupling of ${\bar h}_{00}$ to the Newtonian energy of the two-particle system.   This term can be regarded as a kinetic plus gravitational correction to the mass monopole of the source
\begin{equation}
\sum_a m_a\rightarrow \sum_a m_a \left(1+{1\over 2}{\bf v}^2_a\right) - {G_N m_1 m_2\over |{\bf x}_1-{\bf x}_2|}.
\end{equation}
The second term is a coupling of the graviton to the total mechanical angular momentum of the system, ${\bf L}=\sum_a {\bf x}_a\times m_a {\bf v}_a$.  Both the mass monopole and the angular momentum are conserved at this order in the velocity expansion, and therefore cannot  source on-shell radiation.   Rather they source the long range static gravitational field of the two-body system.    The last term, on the other hand, is a coupling of the moment $\sum_a m_a {\bf x}^i_a {\bf x}^j_a$ to the (linearized) Riemann tensor of the radiation field.    Note that for on-shell radiation $R_{00}= R_{0i0i}=0$, so only the traceless moment
\begin{equation}
Q^{ij} = \sum_a m_a \left({\bf x}^i_a {\bf x}^j_a-{1\over 3}{\bf x}^2_a\delta_{ij}\right)
\end{equation}
is a source of gravitational waves.  It is interesting to note that in order to obtain Eq.~(\ref{eq:LV2}), we needed to include not only the multipole expansion of the couplings of ${\bar h}_{\mu\nu}$ to the point particles, but also diagrams with graviton self-interactions.   This is forced upon us by the power counting rules of the theory.     It would be simply inconsistent to drop the second two graphs in Eq.~(\ref{eq:LV2}) and keep only the first.   Thus even the leading order quadrupole coupling to radiation is sensitive to the non-linearities present in relativistic gravity.   Physically, this is perhaps not too surprising, as the graviton couples to all sources of energy-momentum, including the energy-momentum stored in the gravitational field of the particles themselves.

{\bf Exercise}:   Use the power counting rules to check that 
\begin{eqnarray}
\Diagram{gd\\ hu}
\Diagram{fvA\\ fvA}\sim v^{5/2}, & \Diagram{h &h\\ & gv}\sim {v^{5/2}\over \sqrt{L}},
\end{eqnarray}
so that the second two graphs in Eq.~(\ref{eq:N2LOrad}) each scale as $\sqrt{L} v^{5/2}$.  This is suppressed by $v^2$ relative to the graph in Eq.~(\ref{eq:lv0}).

Working out higher order terms in the EFT is similar, and there is no conceptual problem in carrying out the expansion to any desired order in $v$.   Once we have calculated the terms in $S_{NR}[{\bar h}]$, it is straightforward to compute physical observables such as the gravitational radiation power output.  As discussed in Sec.~\ref{sec:obs}, these can be obtained from
\begin{equation}
\exp[iS_{\mathit eff}(x_a)] = \int D{\bar h}_{\mu\nu}(x) e^{i S_{NR}[{\bar h}]},
\end{equation}
which is simply the sum over diagrams, calculated using the Feynman rules from $S_{NR}[\bar h]$, that have no external graviton lines.    For example, the leading order quadrupole radiation formula follows from
\begin{equation}
\mbox{Im}\,\,\,{\maxis{\Diagram{ms m  gl m ms}}}\Rightarrow P= {G_N\over 5}\langle \stackrel{\ldots}{Q}_{ij} \stackrel{\ldots}{Q}_{ij} \rangle,
\end{equation}
where the brackets denote a time average.   In order to compute the power, one needs to know the time evolution of the moment $Q_{ij}(t)$.     This can be obtained by solving the equations of motion for the ${\bf x}_a(t)$, which follow from the terms in $S_{NR}[{\bar h}]$ with no powers of ${\bar h}_{\mu\nu}$.

\subsection{Finite size effects}
\label{sec:finsize}

At some order in the velocity expansion, the internal structure of the binary star constituents play a role in the dynamics.   In the case of black hole binaries, it is easy to determine the order in $v$ for which this happens.    As discussed in Sec.~\ref{sec:ppEFT}, finite size effects are encoded in non-minimal couplings to the curvature.   Such terms are built out of invariants constructed from the Riemann tensor, the simplest ones being those of Eq.~(\ref{eq:finsize}),
\begin{equation}
S = \cdots + c_E\int d\tau E_{\mu\nu} E^{\mu\nu} + c_B \int d\tau B_{\mu\nu} B^{\mu\nu}+\cdots.
\end{equation}
The coefficients $c_{E,B}$ can be obtained by a matching calculation.  One simply compares some observable in the point particle theory to the analogous quantity in the full theory, adjusting the coefficients to ensure that the two calculations agree.  

For the purposes of calculating the coefficients,  $c_{E,B}$, a convenient observable is the $S$-matrix element for a graviton to scatter off the BH background.   In the EFT, this amplitude is schematically of the form
\begin{equation}
i{\cal A} = \cdots + \Diagram{gd \\gu}\Diagram{{fvA}\\ ![urt]{fvA}{c_{E.B}}}\,\,\,\,\,\,\,\,\,\,+\cdots\sim\cdots + i {c_{E,B}\over m^2_{Pl}}\omega^4+\cdots.
\end{equation}
Here, $\omega$ is the energy of the incoming graviton.   For the point particle EFT to be valid, it must satisfy $r_s\omega\ll 1$.    The factor of ${1/m^2_{Pl}}$ is due to the fact that we are looking at a two graviton process.   We have not shown, for example, the leading order term in the amplitude, due to graviton scattering off the field created by the mass term $-m\int d\tau$ in the point particle action.   Although the leading effect of $c_{E,B}$ is from interference with this term, the above equation predicts that the total scattering cross section contains a term
\begin{equation}
\sigma(\omega)_{EFT}\sim \cdots + {c_{E,B}^2\over m^4_{Pl}}\omega^8+\cdots.
\end{equation}
The only scale in the full theory is the BH radius $r_s$, so the cross section must be of the form
\begin{equation}
\sigma(\omega)_{BH} = r^2_s f(r_s\omega),
\end{equation}
where $f(r_s\omega)$ is a function that for $r_s\omega\ll 1$ can be expanded in powers of $r_s\omega$ (possibly times logs).   Thus we expect to find a term
\begin{equation}
\sigma(\omega)_{BH} \sim \cdots +r^{10}_s\omega^8+\cdots,
\end{equation}
and therefore $c_{E,B}\sim m^2_{Pl}r^5_s$.   Using this scaling, the power counting rules in the non-relativistic limit indicate that the finite size operators first contribute to the dynamics through their effect on the two-body forces, through the diagram
\begin{equation}
\Diagram{![llft]{fvA}{c_{E,B}} \\ fvA}\Diagram{hu\\  hd}\Diagram{fvA \\ fvA}\ \sim L v^{10},
\end{equation}
so is down by ten powers of $v$ relative to the Newton potential.    One concludes that finite size effects are completely irrelevant for binary inspirals.   Actually, this is not completely true, as realistic black holes  usually have non-zero spin, and the inclusion of spin tends to enhance finite size effects~\cite{spin}.   Also, absorption by the black hole horizon, which necessitates the introduction of additional worldline modes in the EFT~\cite{abs} arises at order $v^8$ for non-spinning black holes~\cite{poisson} and at order $v^5$~\cite{spinabs} when the black hole has spin.

\section{Conclusions}
\label{sec:conc}

These lectures presented an introduction to the basic ideas of EFTs and their use in understanding the evolution of coalescing binary stars in the non-relativistic regime.   In treating such systems, one encounters a wide range of physically important length scales, from the internal size of the binary constituents to the gravitational radiation wavelength.   In order to make sense of physics at all these scales, it is extremely convenient to formulate the problem one scale at a time, by constructing a tower of EFTs as outlined in these lectures.

One topic that was not discussed in these lectures is the issue of ultraviolet divergences in the non-relativistic expansion.   Such divergences arise even classically, in the computation of the Feynman diagram that contribute to a given observable.   Physically, the presence of these divergences can be attributed to the singular nature of the point particle limit.   In the EFT approach, these can be handled by the usual regularization and renormalization procedure found in textbooks:    divergences get cutoff using any convenient regulator (e.g. dimensional regularization) and renormalized into the coefficients of operators in the point particle Lagrangian.   Since by construction the point particle EFT has all possible operators consistent with symmetries, all short distance singularities can be removed from the theory.   At high enough orders in the expansion, some of the EFT coefficients are logarithmically renormalized, leading to RG flows that can be exploited to compute terms in the velocity expansion that are logarithmically enhanced.   See~\cite{GnR1} for more details.

We expect that the technology of EFTs should be applicable to other astrophysical sources of relevance to gravitational wave physics.   For example, it is expected that LISA will detect gravitational waves from  the motion of a small object, i.e. a neutron star, around a super-massive black hole.   In this case, there is a small expansion parameter, namely the ratio of the neutron star size to the curvature length of the black hole background.   It seems natural to construct the expansion in this parameter using EFT methods.   

The point-particle EFTs discussed here may also potentially play a role in the problem of tracking the evolution of colliding black holes in numerical relativity. In particular, the EFT  approach to parameterizing black hole internal structure could be useful in addressing the question of how to systematically handle the curvature singularities present in numerical simulations.   More generally, the methods presented in these notes are useful for understanding the dynamics of extended objects interacting with long wavelength fields.   This sort of situation arises in both formal and phenomenological applications.   Work on the ideas outline here is underway.

\section{Acknowledgments}

I would like to thank the students at Session 86 of the Les Houches summer school for their excellent questions, and the organizers, particularly Christophe Grojean, for their hospitality.    I also thank Ira Rothstein and Witek Skiba for  comments on the manuscript, and JiJi Fan for carefully reading parts of the draft.   This work is supported in part by DOE grant DE-FG-02-92ER40704.

\begin{appendix}

\section{Redundant operators}

In general, an effective Lagrangian contains all operators constructed from the light degrees of freedom that are invariant under the symmetries of the low energy theory.   In practice, some operators are redundant, and may be dropped without altering the physical predictions of the EFT.   In fact, operators that vanish by the leading order equations of motion may be omitted from the list of operators appearing in the effective Lagrangian.    Such operators are sometimes called redundant operators.

The basic reason for this is simple.   If an operator vanishes by the leading order equations of motion (i.e., it vanishes ``on-shell''), this means that one can redefine the EFT fields in such a way that the Lagrangian written in terms of the new fields does not contain the redundant operator.   But field redefinitions have no effect on physical observables ($S$-matrix elements) so the Lagrangian without the redundant operator is equivalent to the original Lagrangian.

To see how this works in practice, consider in the EFT for the massless $\pi$ field discussed in Lecture I, the dimension-ten operator
\begin{equation}
{\cal O}_{10}(x) = (\partial^2\pi) F[\pi,\partial_\mu\pi],
\end{equation}
with
\begin{equation}
F=(\partial_\alpha \partial_\beta \pi)  \partial^\alpha\pi \partial^\beta \pi.
\end{equation}
${\cal O}_{10}$ potentially contributes to to the amplitude for $\pi\pi\rightarrow\pi\pi$ scattering at order $(\omega/\Lambda)^6$.    Because to leading order in the $\omega/\Lambda$ expansion the equation of motion for $\pi$ is $\partial^2\pi(x)=0$, this operator actually gives a vanishing contribution to scattering.   This is obvious at tree level, but in fact ${\cal O}_{10}(x)$ may also be dropped from loop diagram contributions to scattering amplitudes as well.

Starting from the Lagrangian including ${\cal O}_{10}(x)$
\begin{equation}
S={1\over 2}\int  d^4 x (\partial_\mu \pi)^2  + \cdots + {c_{10}\over\Lambda^6}\int d^4 x {\cal O}(x)+\cdots,
\end{equation}
define a new field ${\bar\pi}(x)$ by 
\begin{equation}
\pi(x) =  \bar{\pi}(x) + \delta\pi(x).
\end{equation}
Then
\begin{equation}
\int d^4 x (\partial_\mu \pi)^2  = \int d^4 x(\partial_\mu {\bar \pi})^2  - 2 \int d^4 x \partial^2 {\bar \pi} \delta\pi(x) + {\cal O}(\delta\pi^2).  
\end{equation}
So if we choose
\begin{equation}
\delta\pi(x) = {c_{10}\over\Lambda^6} F[{\bar\pi},\partial_\mu{\bar\pi}],
\end{equation}
we see that the shift in the leading order (dimension four) term in the Lagrangian precisely cancels the term with the ${\cal O}_{10}(x)$ in the original Lagrangian.   Note however that the effects of this operator are not completely gone.  For example, inserting the field redefinition into the dimension eight operator ${\cal O}_8(x)=(\partial_\mu\pi\partial^\mu\pi)^2$ generates operators of dimension fourteen and higher
\begin{equation}
(\partial_\mu\pi\partial^\mu\pi)^2 \rightarrow  (\partial_\mu\pi\partial^\mu\pi)^2 + {2 c_{10}\over\Lambda^6} (\partial_\mu\pi\partial^\mu\pi) \partial_\nu \pi \partial^\nu F +\cdots 
\end{equation}   
However, when constructing the original EFT, one must include all possible operators consistent with the symmetries at every order in the $\omega/\Lambda$ expansion.   Thus, the shift in ${\cal O}_8(x)$ due to the field redefinition can be absorbed into the coefficients of a operators that are already present in the theory.   The same is true for the substitution of the field redefinition into any other operator, and we conclude that in terms of the  new variables, the effects of the redundant operator ${\cal O}_{10}(x)$  are completely spurious.

The fact that operators that vanish on-shell may be omitted is very general and very useful.  A systematic discussion is given by Politzer in~\cite{politzer}.   See also~\cite{georgi}.  We used this result in the EFT that describes an extended object coupled to gravity.     The effective theory for the point particle contains the terms
\begin{equation}
c_R\int d\tau R + c_V\int d\tau R_{\mu\nu} {\dot x}^\mu {\dot x}^\nu,
\end{equation}
which are redundant due to the fact that at leading order (ignoring coupling to sources) the Einstein equations imply $R_{\mu\nu}(x)=0$.   Suppose we re-define the metric appearing in the original Einstein-Hilbert Lagrangian, $g_{\mu\nu}\rightarrow g_{\mu\nu} + \delta g_{\mu\nu}$, with
\begin{equation}
\delta g_{\mu\nu}(x)= {1\over 2 m^2_{Pl}} \int d\tau {\delta^4(x-x(\tau))\over\sqrt{g}}\left[-(\xi_R-{1\over 2}\xi_V) g_{\mu\nu} +\xi_V {\dot x}^\mu {\dot x}^\nu\right].
\end{equation}
When plugged into the gravitational action $S_{EH}$ this redefinition induces the shifts 
\begin{equation}
c_{R,V}\rightarrow c_{R,V} + \xi_{R,V}.    
\end{equation}
Thus by adjusting $\xi_{R,V}$ we can make the coefficients $c_{R,V}$ whatever we like, including zero, without affecting the physical predictions of the theory.
\end{appendix}


\begin{thebibliography}{1}

\bibitem{LIGO}
A.~Abramovici {\it et al.}, ``Ligo: The Laser Interferometer Gravitational Wave Observatory,''
  Science {\bf 256}, 325 (1992).

\bibitem{VIRGO}
A.~Giazotto,
  ``THE VIRGO PROJECT: A WIDE BAND ANTENNA FOR GRAVITATIONAL WAVE DETECTION,''
  Nucl.\ Instrum.\ Meth.\ A {\bf 289}, 518 (1990).

\bibitem{LISA}
 K.~Danzmann and A.~Rudiger, ``Lisa Technology - Concept, Status, Prospects,''
  Class.\ Quant.\ Grav.\  {\bf 20}, S1 (2003).

\bibitem{GWrev}
C.~Cutler and K.~S.~Thorne, ``An overview of gravitational-wave sources,''  arXiv:gr-qc/0204090.

\bibitem{kip}
C.~Cutler {\it et al.}, ``The Last three minutes: issues in gravitational wave measurements of coalescing compact binaries,''
  Phys.\ Rev.\ Lett.\  {\bf 70}, 2984 (1993)
  [arXiv:astro-ph/9208005].

\bibitem{PN}
For a review and further references, see L.~Blanchet, ``Gravitational radiation from post-Newtonian sources and inspiralling compact binaries,''
  Living Rev.\ Rel.\  {\bf 5}, 3 (2002)
  [arXiv:gr-qc/0202016].

\bibitem{GnR1}
W.~D.~Goldberger and I.~Z.~Rothstein, ``An effective field theory of gravity for extended objects,''
  Phys.\ Rev.\ D {\bf 73}, 104029 (2006)  [arXiv:hep-th/0409156].   
See also W.~D.~Goldberger and I.~Z.~Rothstein, ``Towers of gravitational theories,''  Gen.\ Rel.\ Grav.\  {\bf 38}, 1537 (2006)  [arXiv:hep-th/0605238].
  

\bibitem{NRQCD}
W.~E.~Caswell and G.~P.~Lepage, ``Effective Lagrangians For Bound State Problems In QED, QCD, And Other Field Theories,'' Phys.\ Lett.\ B {\bf 167}, 437 (1986);

\bibitem{polchinski} J.~Polchinski, ``Effective Field Theory And The Fermi Surface,''  arXiv:hep-th/9210046.  

\bibitem{kaplan} D.~B.~Kaplan, ``Effective field theories,''  arXiv:nucl-th/9506035.

\bibitem{aneesh} A.~V.~Manohar, ``Effective field theories,''  arXiv:hep-ph/9606222.

\bibitem{ira} I.~Z.~Rothstein, ``TASI lectures on effective field theories,''
  arXiv:hep-ph/0308266.  

\bibitem{wilson} For a review, see K.~G.~Wilson and J.~B.~Kogut, ``The Renormalization group and the epsilon expansion,''  Phys.\ Rept.\  {\bf 12}, 75 (1974).
  
 
  
\bibitem{ac}
 T.~Appelquist and J.~Carazzone, ``Infrared Singularities And Massive Fields,''
  Phys.\ Rev.\ D {\bf 11}, 2856 (1975).
  
\bibitem{weak} An extensive review of EFT methods in weak processes is
A.~J.~Buras, ``Weak Hamiltonian, CP violation and rare decays,''  arXiv:hep-ph/9806471.

\bibitem{thresh} 
 S.~Weinberg, ``Effective Gauge Theories,''
  Phys.\ Lett.\ B {\bf 91}, 51 (1980);
  L.~J.~Hall,
  ``Grand Unification Of Effective Gauge Theories,''
  Nucl.\ Phys.\ B {\bf 178}, 75 (1981).
  
\bibitem{chi1} 
S.~Weinberg, ``Phenomenological Lagrangians,''  PhysicaA {\bf 96}, 327 (1979).

\bibitem{chi2}
 J.~Gasser and H.~Leutwyler, ``Chiral Perturbation Theory To One Loop,''   Annals Phys.\  {\bf 158}, 142 (1984);  
  J.~Gasser and H.~Leutwyler, ``Chiral Perturbation Theory: Expansions In The Mass Of The Strange Quark,''   Nucl.\ Phys.\ B {\bf 250}, 465 (1985).
  
\bibitem{donoghue}
J.~F.~Donoghue, ``Introduction to the Effective Field Theory Description of Gravity,'' arXiv:gr-qc/9512024;


\bibitem{burgess}
C.~P.~Burgess, ``Quantum gravity in everyday life: General relativity as an effective field theory,''
arXiv:gr-qc/0311082;
  C.~P.~Burgess, ``Introduction to Effective Field Theory,''  arXiv:hep-th/0701053.




\bibitem{EWEFT}  W.~Buchmuller and D.~Wyler, ``Effective Lagrangian Analysis Of New Interactions And Flavor Conservation,''  Nucl.\ Phys.\ B {\bf 268}, 621 (1986);  
B.~Grinstein and M.~B.~Wise, ``Operator analysis for precision electroweak physics,''  Phys.\ Lett.\ B {\bf 265}, 326 (1991).  

\bibitem{skiba} Z.~Han and W.~Skiba, ``Effective theory analysis of precision electroweak data,''  Phys.\ Rev.\ D {\bf 71}, 075009 (2005)  [arXiv:hep-ph/0412166].

\bibitem{LMR}
M.~E.~Luke, A.~V.~Manohar and I.~Z.~Rothstein, ``Renormalization group scaling in nonrelativistic QCD,'' Phys.\ Rev.\ D {\bf 61}, 074025 (2000)
[arXiv:hep-ph/9910209]. 

\bibitem{QNM}
K.~D.~Kokkotas and B.~G.~Schmidt, ``Quasi-normal modes of stars and black holes,''
  Living Rev.\ Rel.\  {\bf 2}, 2 (1999)
  [arXiv:gr-qc/9909058];
H.~P.~Nollert,
  Class.\ Quant.\ Grav.\  {\bf 16}, R159 (1999).


\bibitem{spin}
R.~A.~Porto, ``Post-Newtonian corrections to the motion of spinning bodies in NRGR,''
  Phys.\ Rev.\ D {\bf 73}, 104031 (2006)
  [arXiv:gr-qc/0511061]; 
R.~A.~Porto and I.~Z.~Rothstein, ``The hyperfine Einstein-Infeld-Hoffmann potential,''
  Phys.\ Rev.\ Lett.\  {\bf 97}, 021101 (2006)
  [arXiv:gr-qc/0604099].


\bibitem{veltman}
M. Veltman, in \emph{Methods in Field Theory, Proceedings of the Les Houches Summer School, 1975,} 
eds. R. Balian and J. Zinn-Justin, North Holland, 1976.







\bibitem{background}
B.~S.~Dewitt, ``Quantum Theory Of Gravity. Ii. The Manifestly Covariant Theory,''
Phys.\ Rev.\  {\bf 162}, 1195 (1967);
L.~F.~Abbott, ``The Background Field Method Beyond One Loop,''
Nucl.\ Phys.\ B {\bf 185}, 189 (1981).



\bibitem{EIH}
A.~Einstein, L.~Infeld and B.~Hoffmann, ``The Gravitational Equations And The Problem Of Motion,''
Annals Math.\  {\bf 39}, 65 (1938).  


\bibitem{QCDmult}
B.~Grinstein and I.~Z.~Rothstein,
Phys.\ Rev.\ D {\bf 57}, 78 (1998)
[arXiv:hep-ph/9703298].

\bibitem{abs}
 W.~D.~Goldberger and I.~Z.~Rothstein, ``Dissipative effects in the worldline approach to black hole dynamics,''  Phys.\ Rev.\ D {\bf 73}, 104030 (2006)  [arXiv:hep-th/0511133].
  
\bibitem{poisson}
E.~Poisson and M.~Sasaki, ``Gravitational radiation from a particle in circular orbit around a black hole. 5: Black hole absorption and tail corrections,''
  Phys.\ Rev.\ D {\bf 51}, 5753 (1995)
  [arXiv:gr-qc/9412027].

 \bibitem{spinabs}
 H.~Tagoshi, S.~Mano and E.~Takasugi, ``Post-Newtonian expansion of gravitational waves from a particle in circular orbits around a rotating black hole: Effects of black hole absorption,''
  Prog.\ Theor.\ Phys.\  {\bf 98}, 829 (1997)
  [arXiv:gr-qc/9711072].

\bibitem{politzer}
H.~D.~Politzer, ``Power Corrections At Short Distances,''
  Nucl.\ Phys.\ B {\bf 172}, 349 (1980).

\bibitem{georgi}
 H.~Georgi, ``On-Shell Effective Field Theory,''  Nucl.\ Phys.\ B {\bf 361}, 339 (1991).





\end{thebibliography}
\end{document}